\documentclass[useAMS,usenatbib]{mn2e}

\usepackage{graphicx}
\usepackage{lscape}
\usepackage{amssymb}

\newcommand{\Sun}{_{\sun}}

\newcommand{\damp}{_{\mathrm{damp}}}

\newcommand{\degree}{^o}

\newcommand{\K}{\,\textrm{K}}
\newcommand{\KeV}{\,\textrm{keV}}
\newcommand{\Kpc}{\,\textrm{kpc}}
\newcommand{\Mpc}{\,\textrm{Mpc}}

\newcommand{\Myr}{\,\textrm{Myr}}
\newcommand{\Gyr}{\,\textrm{Gyr}}

\newcommand{\gccm}{\,\textrm{g}\,\textrm{cm}^{-3}}

\title[Fast gas sloshing simulations]{Fast simulations of gas sloshing and cold front formation}

\author[E. Roediger \& J. ZuHone]%
{E. Roediger$^{1}$\thanks{E-mail: e.roediger@jacobs-university.de},
J. A. ZuHone$^{2}$\\
$^{1}$Jacobs University Bremen, PO Box 750 561, 28725 Bremen, Germany\\
$^{2}$ NASA/Goddard Space Flight Center, 8800 Greenbelt Rd., Code 662, Greenbelt, MD 20771, USA}

\begin{document}

\date{Accepted 1988 December 15. Received 1988 December 14; in original form 1988 October 11}

\pagerange{\pageref{firstpage}--\pageref{lastpage}} \pubyear{2011}

\maketitle

\label{firstpage}

\begin{abstract}
We present a simplified and fast method for simulating minor mergers between galaxy clusters. Instead of following the evolution of the dark matter halos directly by the N-body method, we employ a rigid potential approximation for both clusters. The simulations are run in the rest frame of the more massive cluster and account for the resulting inertial accelerations in an optimised way. We test the reliability of this method for studies of minor merger induced gas sloshing by performing a one-to-one comparison between our simulations and hydro+N-body ones. We find that the rigid potential approximation reproduces the sloshing-related features well except for two artefacts: the temperature just outside the cold fronts is slightly over-predicted, and the outward motion of the cold fronts is delayed by  typically 200 Myr. We discuss reasons for both artefacts. 
\end{abstract}




\section{Introduction}
%
During the last decade, high resolution X-ray observations have found a wealth of structure in the intra-cluster medium (ICM) of galaxy clusters, among them cold fronts (see review by \citealt{Markevitch2007}). These structures reveal themselves as sharp discontinuities in X-ray brightness accompanied by a jump in temperature, where the brighter side is the cooler one. One variety of cold fronts (CFs) was soon understood to be the contact discontinuity between the gaseous atmospheres of merging clusters (e.g.~A2142: \citealt{Markevitch2000}; A3667: \citealt{Vikhlinin2001}; and the bullet cluster 1E 0657-56: \citealt{Markevitch2002}). 

A second class of CFs was found to form arcs around the cool cores of apparently relaxed clusters (%
e.g.~RX J1720.1+2638: \citealt{Mazzotta2001,Mazzotta2008,Owers2009hifid}; 
MS1455.0+2232: \citealt{Mazzotta2008,Owers2009hifid}; 2A0335+096 \citealt{Mazzotta2003,Sanders2009_2a}; 
A2029: \citealt{Clarke2004,Million2009sample}; 
A1795: \citealt{Markevitch2001,Bourdin2008}; 
Perseus: \citealt{Churazov2003,Sanders2005perseus}; 
A496: \citealt{Dupke2007,Ghizzardi2010}; 
Virgo: \citealt{Simionescu2010};
Centaurus: \citealt{Fabian2005centaurus,Sanders2006centaurus}). 
\citet{Markevitch2001} suggested that this variety of CFs forms due to sloshing of the cool central gas within the central cluster potential, where the sloshing is initially triggered by a minor merger event. 
Using  hydro+N-body simulations, \citet{Ascasibar2006} (AM06 afterwards) have  shown that the sloshing scenario reproduces the morphology of observed CFs. 

In recent years, more and more sloshing CFs have been reported (e.g.~\citealt{Owers2009hifid,Ghizzardi2010}). In principle, the properties of the CFs contain information about the merger history of the clusters, as has been shown for the Virgo cluster (\citealt{Roediger2011}, R11a hereafter),  A496 (\citealt{Roediger2011a496}, R11b hereafter), and  RXJ1347.5-1145 (\citealt{Johnson2011}). However, disentangling the merger history of each cluster from the  CF properties requires a set of dedicated simulations for each cluster. Doing this with full hydro+N-body simulations is computationally expensive. The same is true if the influence of more time-consuming physics like viscosity (\citealt{ZuHone2010}), magnetic fields (\citealt{ZuHone2011}) or thermal conduction (ZuHone et al. 2011, in prep.) on sloshing CFs is studied.  A reasonable simplification that speeds up the simulations considerably would be very useful. The most expensive part of such simulations is the self-gravity of the gas and dark matter (DM) particles. In most cases, the self-gravity of the gas can be neglected, because the Jeans length is about 1 Mpc. The simulations  speed up substantially when the DM halos of the main cluster and the subcluster are  approximated as rigid potentials (RPs). However, AM06 have shown that even in a minor merger the central part of the main cluster moves significantly w.r.t.~the overall  cluster potential, thus this effect cannot be neglected. \citet{ZuHone2010} (Z10 hereafter) suggested that the rigid potential approximation can be used if additionally a point-mass-like approximation of  the motion of both clusters is taken into account, including the effects of inertial acceleration.

Here we improve this rigid potential approximation and demonstrate the reliability of the simplified simulations for the application to gas sloshing by comparing them to hydro+N-body simulations. This rigid potential approximation has already been applied successfully in sloshing simulations for the Virgo cluster (R11a) and for A496 (R11b).

\section{Method}
%
\subsection{Test setup}  \label{sec:clustermodel}
We consider the following scenario: The ICM in a massive spherical galaxy cluster (the main cluster) is initially in hydrostatic equilibrium. A spherical, less massive, gas free galaxy cluster (the subcluster) passes through the main cluster. The gravitational impact of the subcluster  initiates sloshing of the ICM in the main cluster core and subsequent cold front formation (see AM06 for a detailed description of the dynamics). 

Hydro+N-body simulations of this scenario have been performed by AM06 and Z10. Our aim here is to investigate to what extent simulations with a rigid potential approximation (RPA) for the DM content of, both, the main cluster and the subcluster can reproduce the resulting CF structures in terms of morphology, orientation, size, temperature and density distribution. As the reference, we use the hydro+N-body simulations of Z10. Thus, we follow their lead and tailor our initial models to match theirs. The main cluster model is based on a Hernquist potential (\citealt{Hernquist1990})  with a scale radius, $a$. The temperature profile is described by the phenomenological function
\begin{equation}
T(r)= \frac{T_0}{1+ r/a} \, \frac{c+ r/a_c}{1+ r/a_c},
\end{equation}
where $T_0$ is a measure for the overall cluster temperature, $c$ describes the depth of the central density drop, and $a_c$ characterises the radius of this drop (see also AM06). The corresponding density profile resulting from hydrostatic equilibrium is (AM06)
\begin{eqnarray}
\rho(r) &=& \rho_0 \left(  1+\frac{r}{a_c}  \right)    \left(  1+\frac{r}{ca_c}  \right)^\alpha    \left(  1+\frac{r}{a}  \right)^\beta \\
\textrm{with}&& \alpha = -1-n\frac{c-1}{c-a/a_c}\,\,\textrm{and}\,\,\beta=1-n\frac{1-a/a_c}{c-a/a_c}. \nonumber
\end{eqnarray}
%
In our setup, we initialise the main cluster with these density and temperature profiles, choosing the parameters such that they fit the corresponding hydro+N-body simulation (see Table~\ref{tab:models}). From these profiles, we derive the gravitational potential of the main cluster assuming hydrostatic equilibrium. 

As in Z10, the subcluster is initialised as a pure DM structure, where the DM mass distribution is described by a Hernquist profile (\citealt{Hernquist1990}).

\subsection{The rigid potential approximation} \label{sec:inertframe}
Our simulations are run in the rest frame of the main cluster. The gas dynamics is described by the hydrodynamical equations. Additionally, the ICM is subject to the gravitational acceleration due to the main cluster and the subcluster.

\subsubsection{Orbit of the subcluster}
We assume the orbit of the subcluster to be the orbit of a test particle free-falling through the main cluster. In the course of the simulation, the potential of the subcluster is shifted through the main cluster along this orbit.  This approach does not include  dynamical friction, which will slow down the subcluster after pericentre passage (see Sect.~\ref{sec:orbit} for a comparison).  However, the sloshing is triggered mainly during the pericentre passage, and thus our results do not suffer from this discrepancy. Given that our test particle orbit is bound to predict increasingly  wrong subcluster positions after the first pericentre passage, we stop  our simulations well before a second pericentre passage. 

We construct orbits that are comparable to the ones in the hydro+N-body simulations in orientation, pericentre distance, and velocity history prior to pericentre passage. We start our simulations $1\Gyr$ prior to the pericentre passage of the subcluster, which we set to occur at $t=0$. We reran our fiducial simulation with an initialisation time of $0.5\Gyr$ prior to pericentre passage and found that our results are not sensitive to the choice of the initialisation time.

The subcluster orbit is in the $xy$-plane of the computational grid. For the sake of a short notation, we identify the $+y$-direction as "north" (N), the $-y$-direction as "south" (S), the $+x$-direction as "west" (W), and the $-x$-direction as "east" (E). The subcluster will start W of the main cluster core, has its closest approach to the main cluster core in the NE and moves away towards the SE.

\subsubsection{Basic inertial frame correction} \label{sec:inertframe_basic}
As the rest frame of the main cluster is not an inertial frame, the ICM in this frame is subject to a pseudo-acceleration due to the attraction of the main cluster core towards the approaching subcluster. Z10 used the most simple approximation to account for this: they assumed that the main cluster responds to the gravity of the subcluster as a whole, like a rigid body. Consequently, they calculate the inertial acceleration felt by the main cluster centre due to the subcluster and add this pseudo-acceleration to  all of the ICM. 

\subsubsection{Improved inertial frame correction}  \label{sec:method_damping}
This basic inertial frame correction is a reasonable approximation for the central region of the main cluster, but it is wrong for the outer parts of the cluster. It will lead to unrealistic flows in the outer parts of the main cluster. For smaller pericentre distances, during pericentre passage of the subcluster the inertial acceleration can be large and even produce supersonic motions in the cluster outskirts. These unrealistic flows can influence the resulting CFs at later stages. Hence, we propose to apply the pseudo-acceleration only to the central region of the main cluster inside a characteristic radius, $R\damp$, and dampen it outside this radius exponentially over a length scale, $L\damp$. Thus, instead of adding the same inertial frame acceleration at every position in the cluster, we multiply it with a radius-dependent function,
\begin{equation}
W(r)=\left\{
\begin{array}{ll}
1  &  \mathrm{if}\;r\le R\damp    \\
\exp(-\frac{r-R\damp}{L\damp})  &     \mathrm{else}    
\end{array}
 \right.,
\end{equation}
where $r$ is the distance to the main cluster centre. 

The choice of $R\damp$ is motivated by the sphere of influence of the subcluster: If the subcluster passes the main cluster centre at a pericentre distance larger than its own size, it attracts the main cluster core only slightly. If the subcluster passes the main cluster core at a small distance, it can attract at most a region comparable to its own  size, but not beyond that.  Hence,  $R\damp$ should be comparable to the size of the subcluster, i.e., about twice its characteristic scale length. We test several combinations of $(R\damp,L\damp)$.

\subsection{Code}
All simulations are run with the FLASH code (version 3.2, \citealt{Dubey2009}).  FLASH is a modular block-structured AMR code, parallelised using the Message Passing Interface (MPI) library. It solves the Riemann problem on a Cartesian grid using the Piecewise-Parabolic Method (PPM). The simulations are performed in 3D and all boundaries are reflecting. We use a simulation grid of size $4\times 3\times 3\Mpc^3$, which is large enough to prevent reflected waves reaching the central region of interest during our simulation time.  We resolve the   inner 50 kpc with 2 kpc, the inner 130 kpc with 4 kpc, the inner 260 kpc with 8 kpc and enforce decreasing resolution with increasing radius from the cluster centre to optimise computational costs. We have performed resolution tests here and also in R11a,b and found our results to be independent of resolution.

\section{Results}
%
\begin{table}
\caption{Summary of model parameters. See Sect.~\ref{sec:clustermodel} for details.}
\begin{center}
\begin{tabular}{llll}
\hline
\multicolumn{2}{l}{\textbf{merger characteristics}}  &  &  \\
mass ratio                        & 20  & 5      & 2     \\
impact parameter  & 200 kpc &  500 kpc & 500 kpc \\
pericentre distance  &  62.5  kpc   & 150 kpc &  135 kpc\\
  &  &  &  \\
\textbf{main cluster}   &  &  &  \\
    $\rho_0 /(10^{-25}\gccm)$      & $3.75$ & $2.9$ &  $2.76$ \\
     $a/\Kpc$             & 615 & 623 &  615\\
      $n$                        & 5 & 5 &  5\\
       $T_0 / (10^7\K)$                     & $15.8$  & $13.8 $ & 11.27 \\
     $a_c$                        & 61.5 & 62.3 &  61.5\\
      $c$                        & 0.1585 & 0.169 & 0.1557 \\
        &  &  &  \\
\textbf{subcluster}    &  &  &  \\      
mass  ($10^{14}M\Sun$)   & $0.714$  & $2.5$ & 5 \\
scale radius  (kpc) & 220  &      350 &  416\\
   &  &  &  \\
    \textbf{damping}    & none,          & none,            &  none, \\      
 $(R\damp/\Kpc,$       & (500, 300)  & (800, 300),   &  (950, 400) \\
 $L\damp/\Kpc) $       &                    & (800, 100),   &  \\
                                   &                    & (500, 300)    &  \\

   \hline
\end{tabular}
\end{center}
\label{tab:models}
\end{table}%
%
We present  simulations  of three merger scenarios which cover a range in mass ratios from 2 to 20. The complete model parameters for each run are listed in Table~\ref{tab:models}.

In the following subsections we give a detailed comparison of the RP simulations to the hydro+N-body ones. In order to aid the reader in smoothly following our analysis, we state our main finding already here: The RPA is able to reproduce the hydro+N-body simulations very well except for two systematic differences, of which one can be fully corrected for, and the other partially:
\begin{itemize}
\item After the onset of sloshing, the RP simulations lag behind the hydro+N-body ones by 200 to 250 Myr, depending on the cluster mass ratio. This lag can be corrected for by delaying the hydro+N-body results by the appropriate lag in the comparison.%
\footnote{In a correct manner, the results of the RPA should be brought forward instead of delaying the more accurate hydro+N-body ones. However,  there are several RP runs for each merger case and only one hydro+N-body run. For the sake of simplicity and clarity we decide to apply the time shift to the hydro+N-body run and trust the reader to remember this footnote.} 
\item The RPA generally over-estimates the temperature just outside the CFs. This disagreement can be attenuated by the damping discussed above, but not avoided completely.
\end{itemize}

\subsection{Fiducial case: mass ratio 5, moderate damping}
Our fiducial case is the intermediate merger with a mass ratio of 5 between the clusters. We simulate this case with four different settings for the inertial frame correction as listed in Table~\ref{tab:models}: one without large-scale damping, two moderate damping settings and a strong one. The best results are achieved for moderate damping with $(R\damp/\Kpc,L\damp/\Kpc)=(800, 300)$. This case will be described first. The effect of  the damping will be summarised in Sect.~\ref{sec:fiducial_damping}.

\subsubsection{Morphology and orientation of cool spiral}
%
\begin{figure*}
\rotatebox[origin=l]{90}{\phantom{bitspace}hydro+Nbody}
\includegraphics[trim=0   0 300     0,clip,height=0.2\textheight]{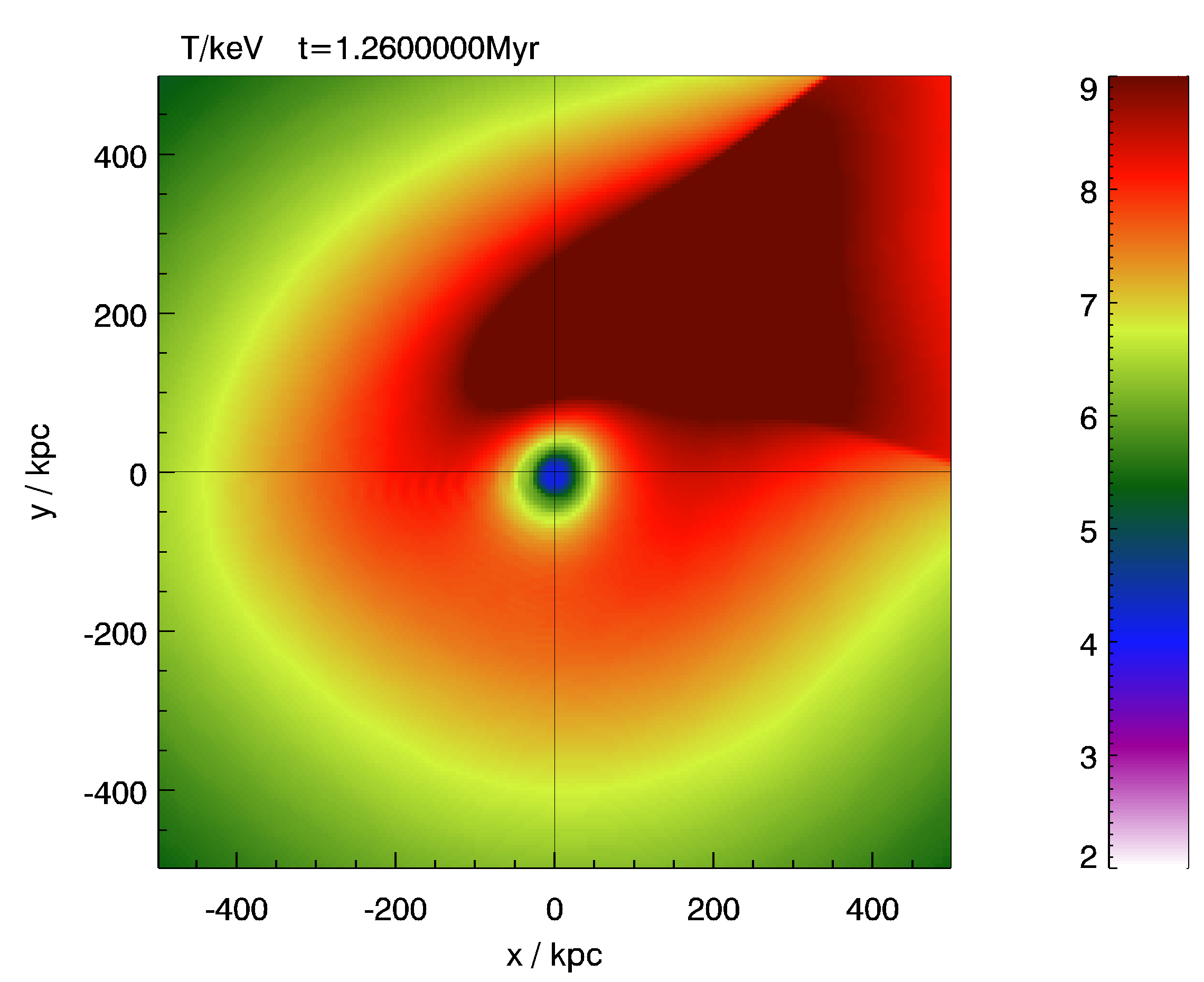}
\includegraphics[trim=300   0 300     0,clip,height=0.2\textheight]{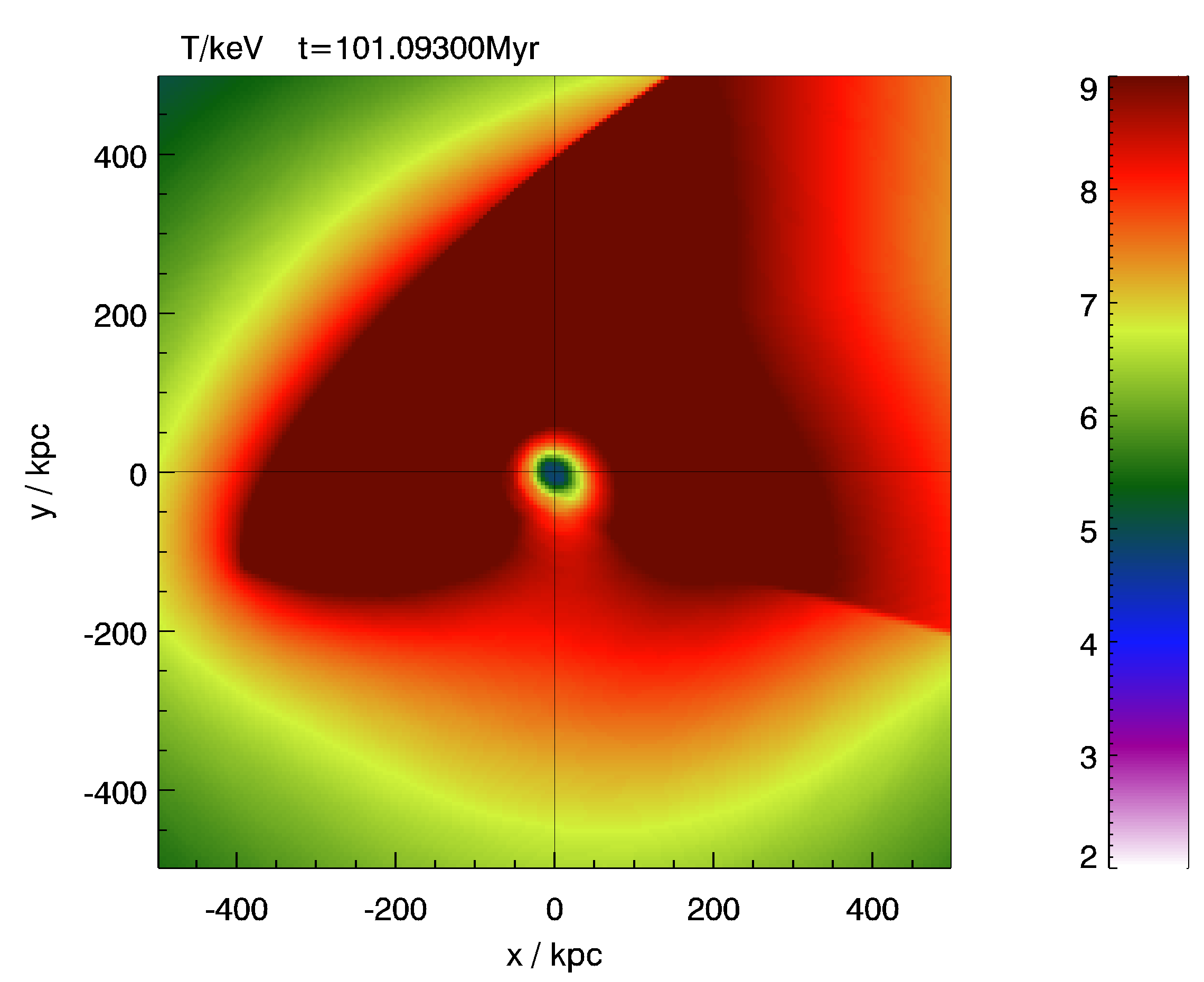}
\includegraphics[trim=300   0 0     0,clip,height=0.2\textheight]{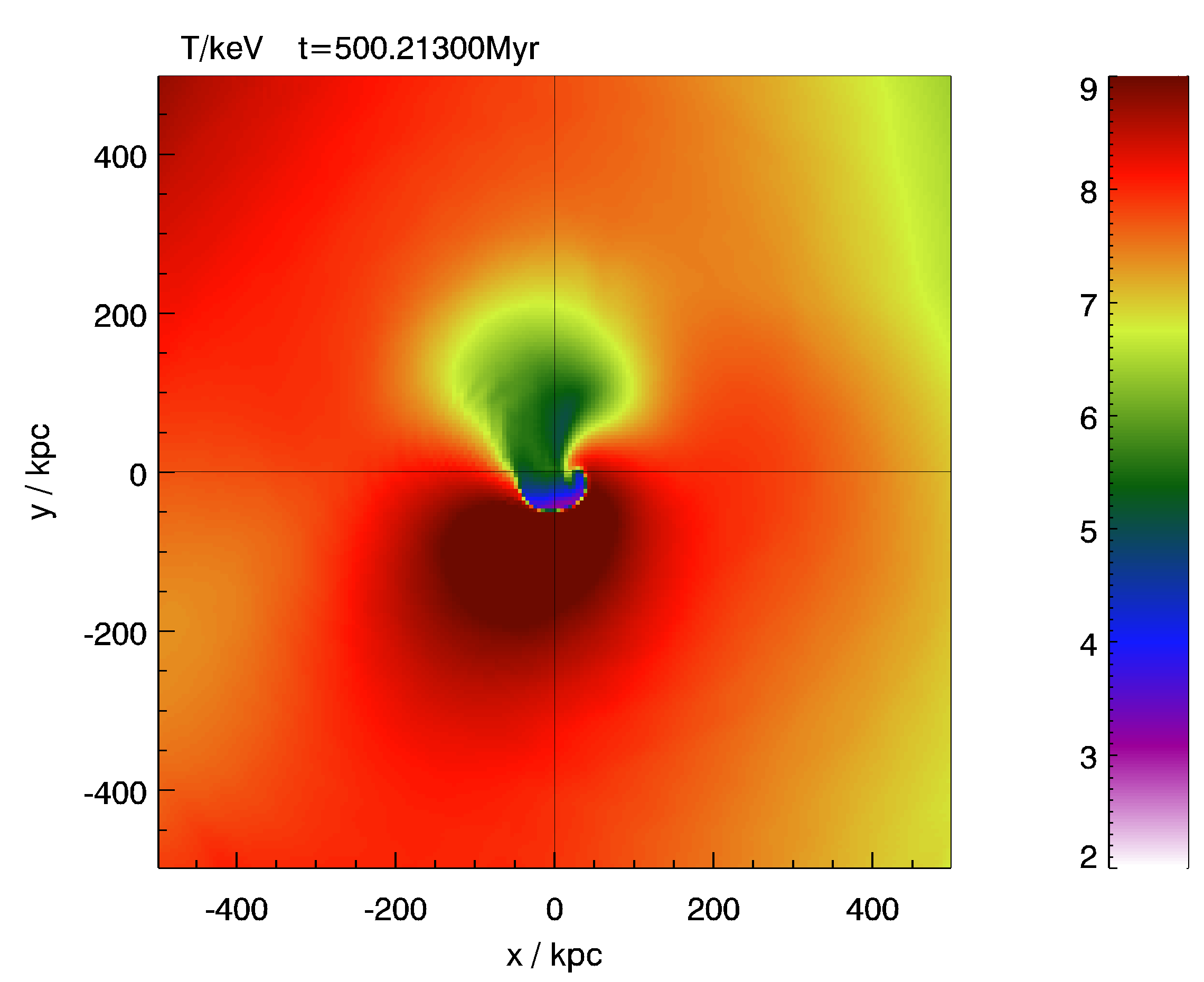}
\newline
\rotatebox[origin=l]{90}{\phantom{bitspace}rigid potential}
\includegraphics[trim=0   0 300     0,clip,height=0.2\textheight]{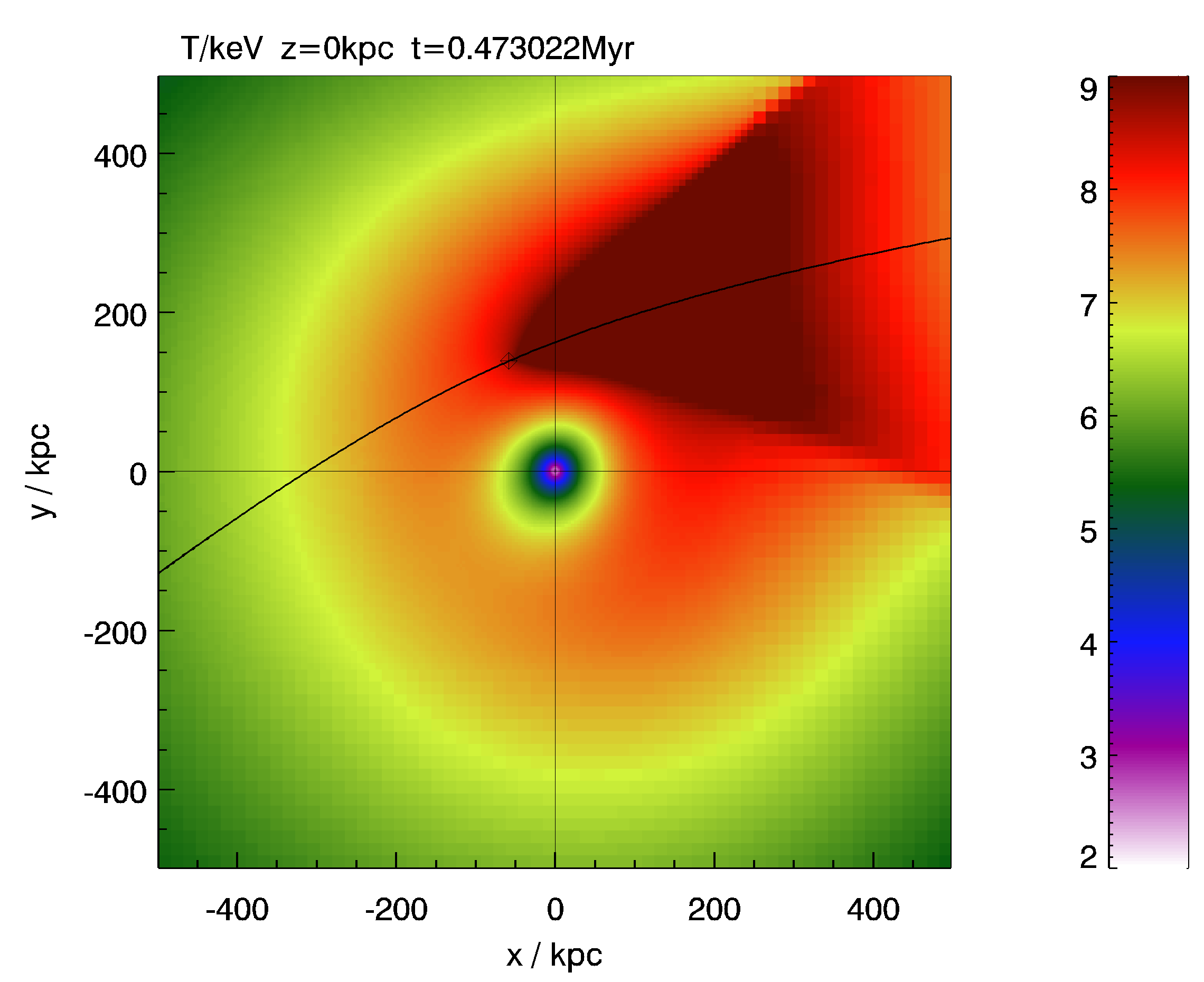}
\includegraphics[trim=300   0 300     0,clip,height=0.2\textheight]{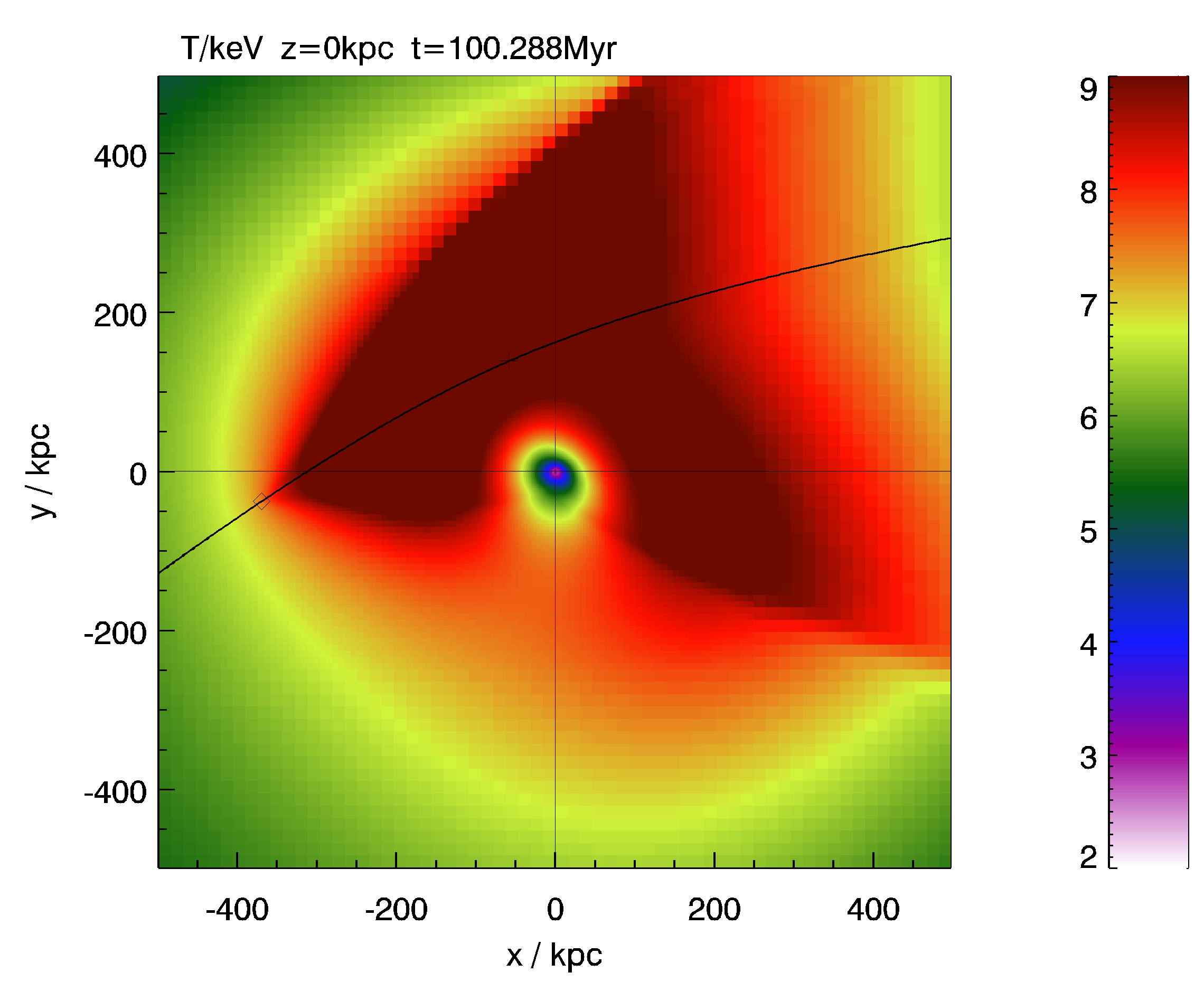}
\includegraphics[trim=300   0 0     0,clip,height=0.2\textheight]{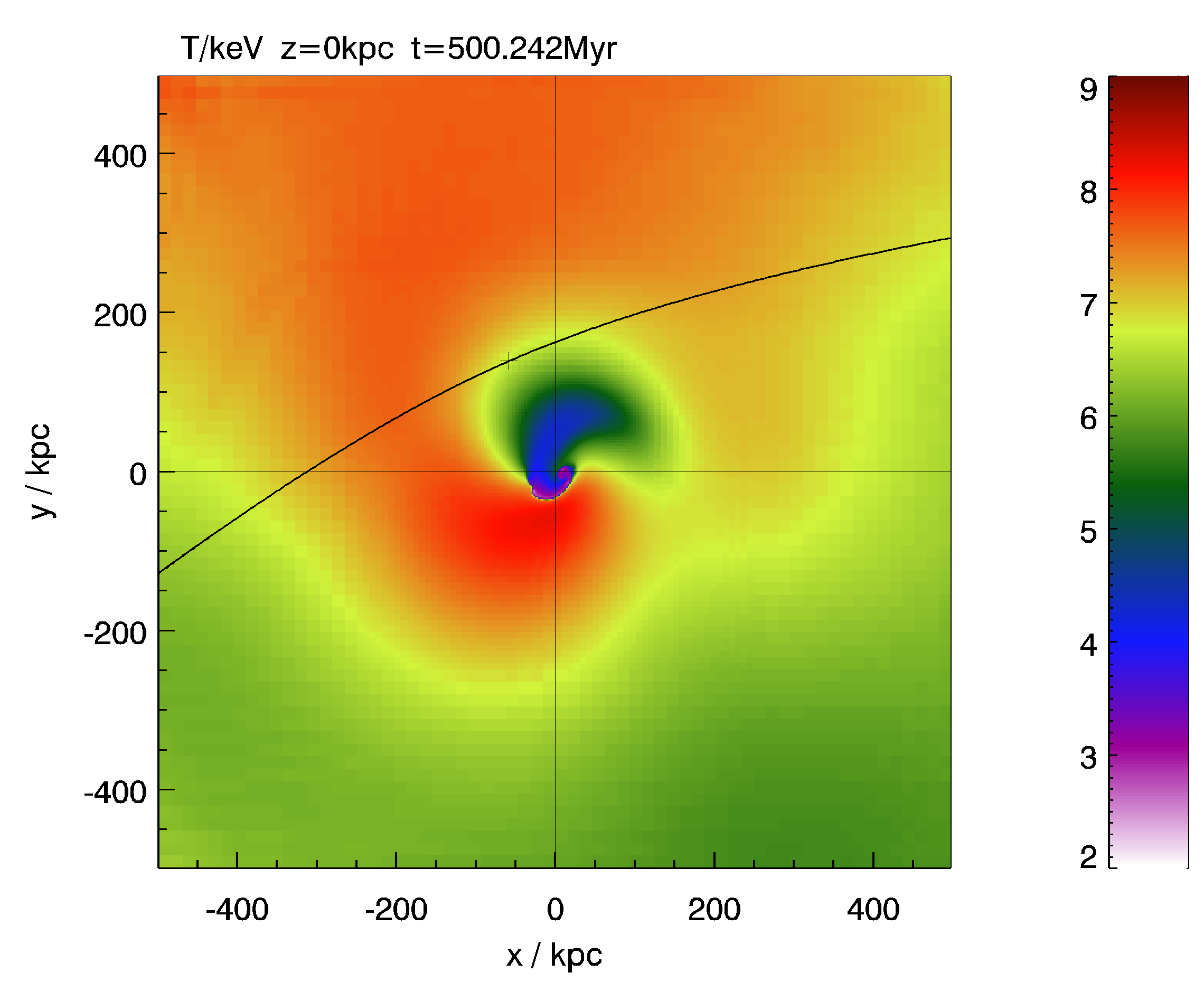}
\newline
\newline
\newline
\rotatebox[origin=l]{90}{\phantom{bitspace}hydro+Nbody}
\includegraphics[trim=0   0 300     0,clip,height=0.2\textheight]{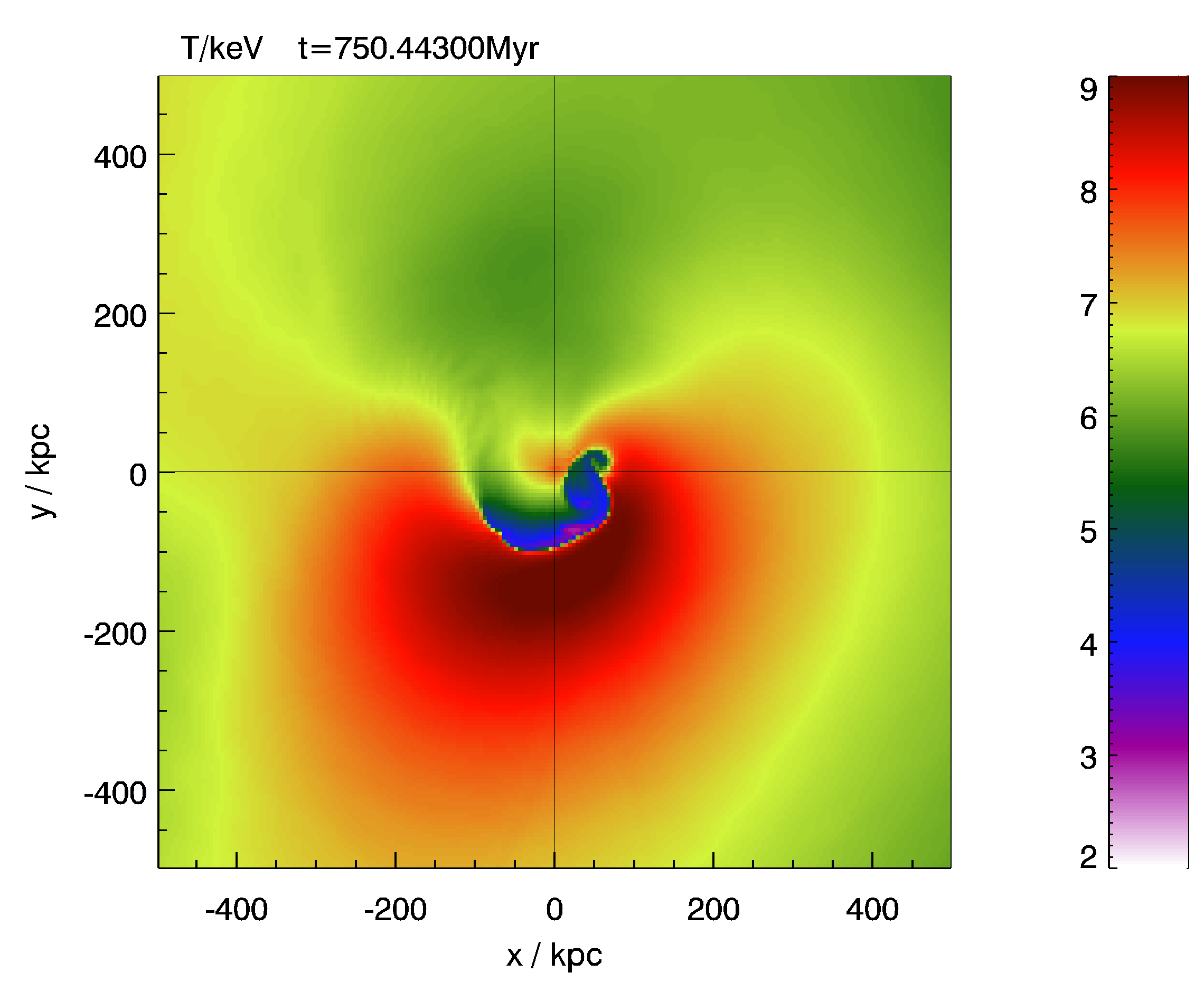}
\includegraphics[trim=300   0 300     0,clip,height=0.2\textheight]{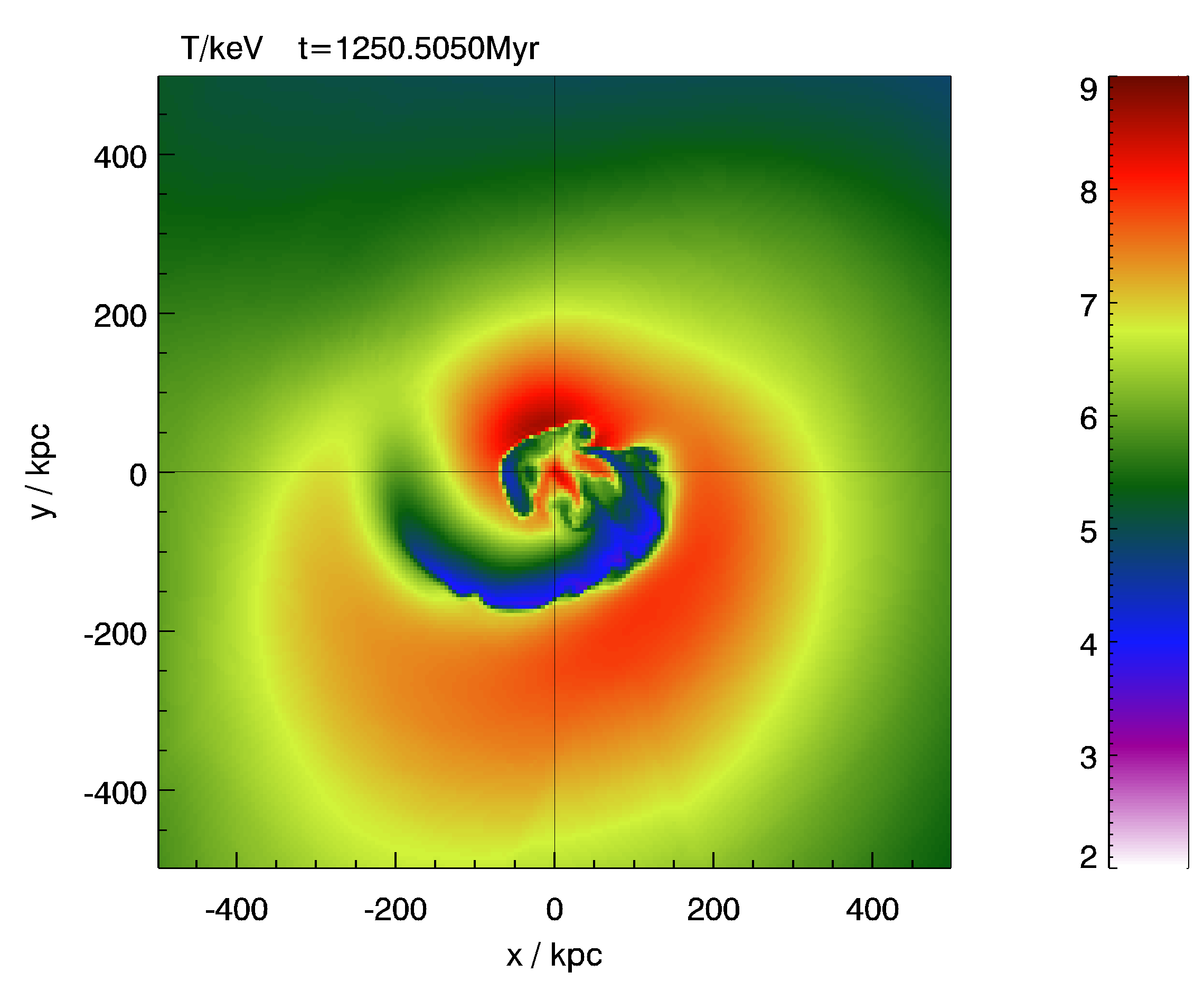}
\includegraphics[trim=300   0 0     0,clip,height=0.2\textheight]{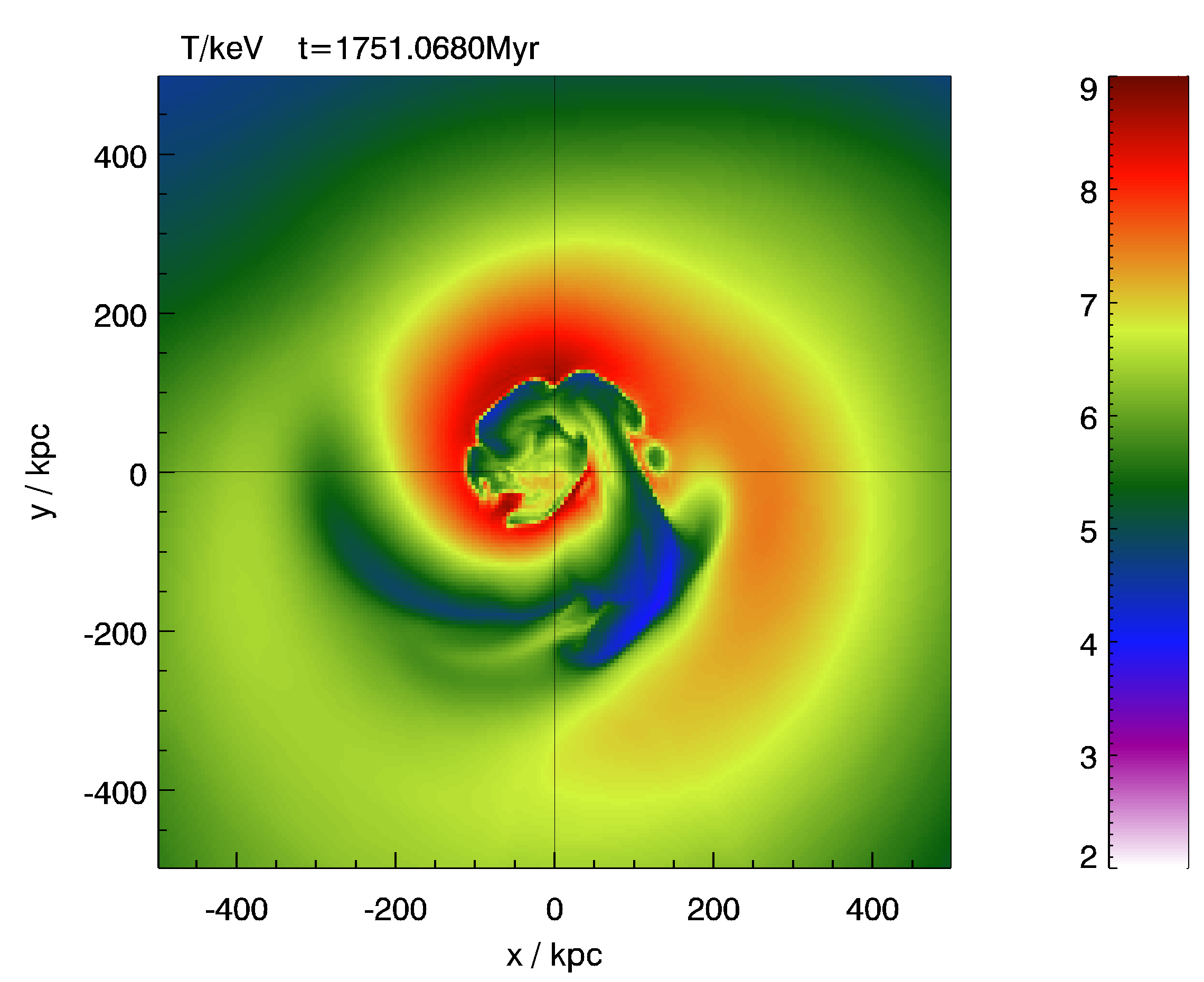}
\newline
\rotatebox[origin=l]{90}{\phantom{bitspace}rigid potential}
\includegraphics[trim=0   0 300     0,clip,height=0.2\textheight]{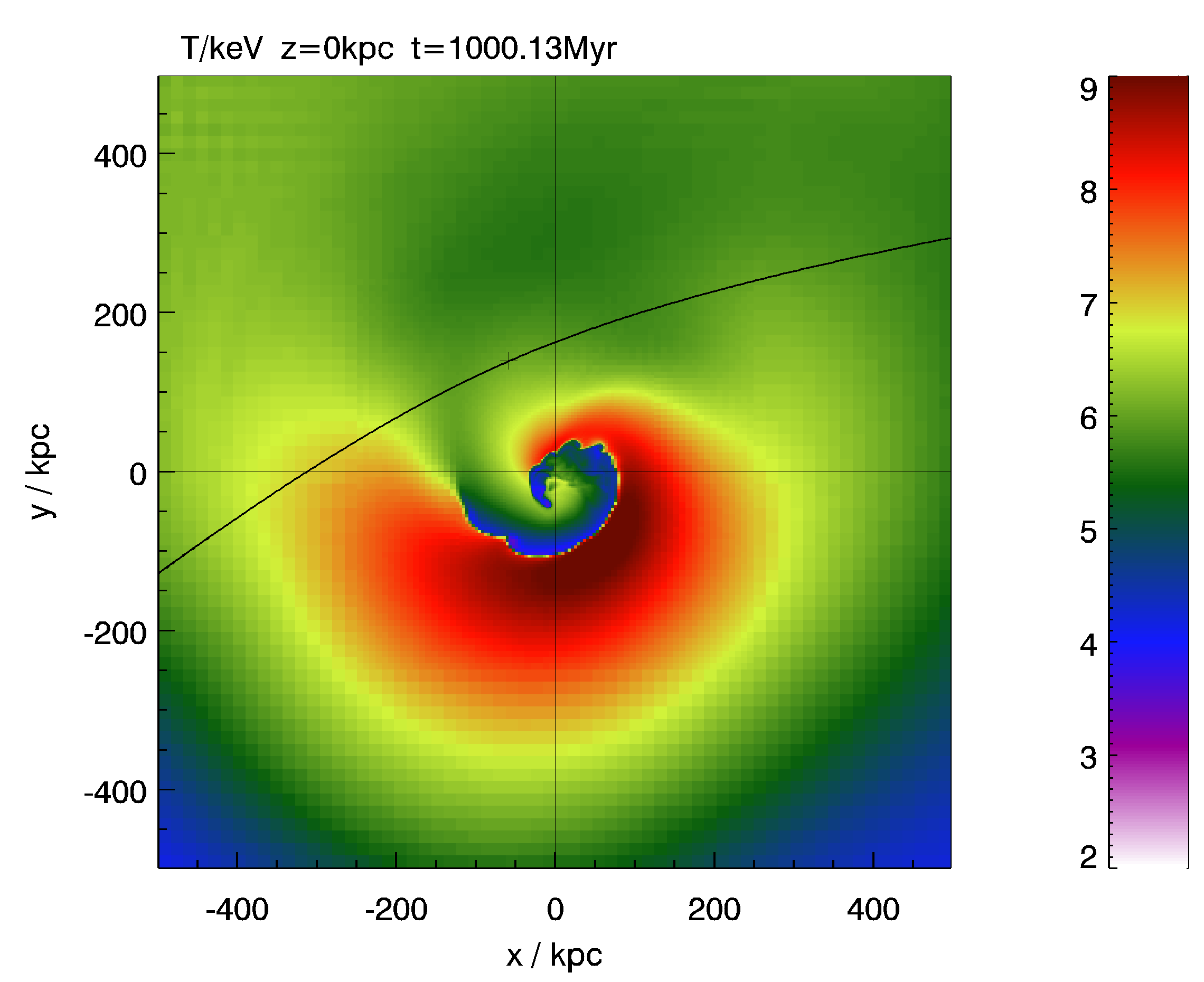}
\includegraphics[trim=300   0 300     0,clip,height=0.2\textheight]{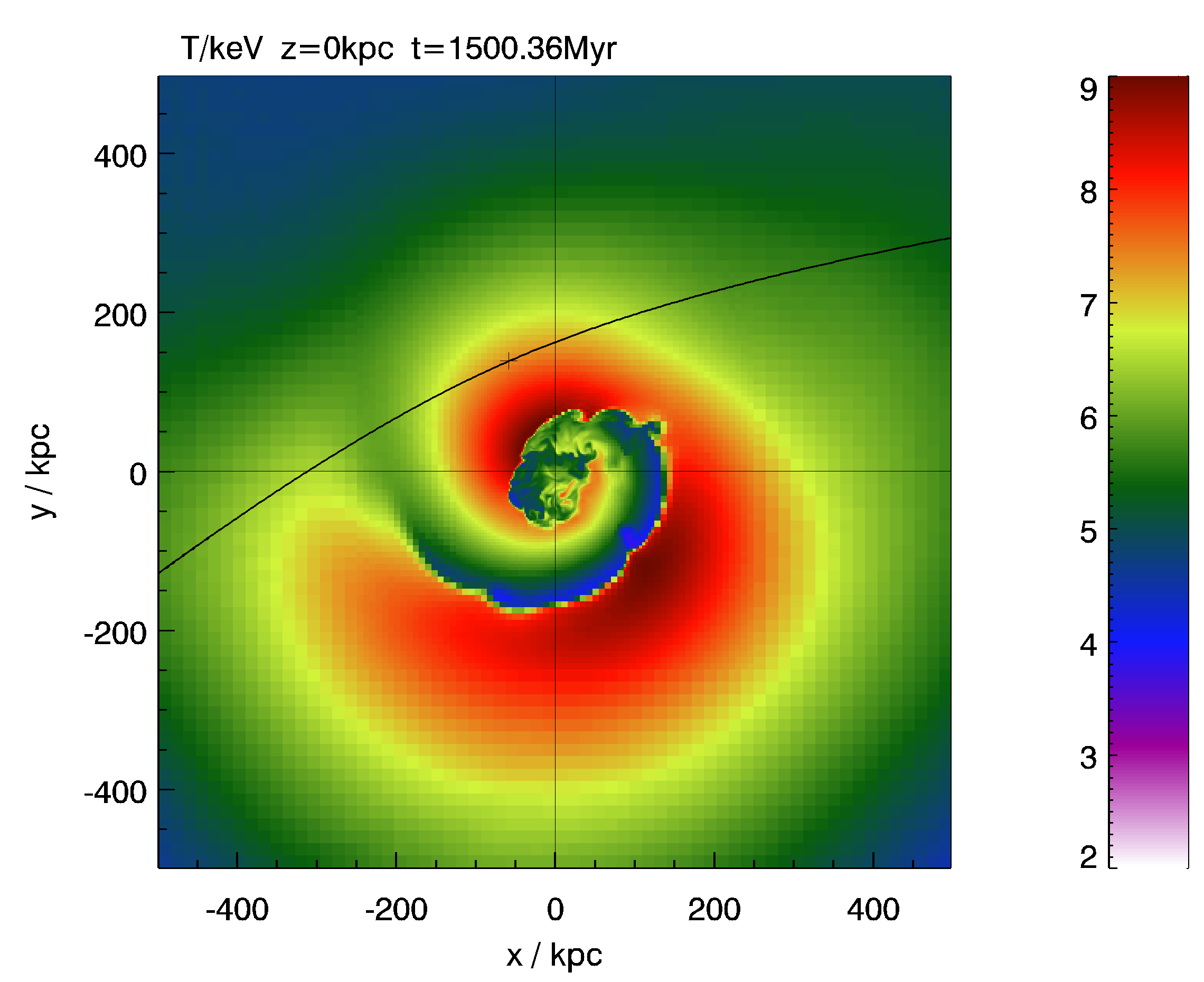}
\includegraphics[trim=300   0 0     0,clip,height=0.2\textheight]{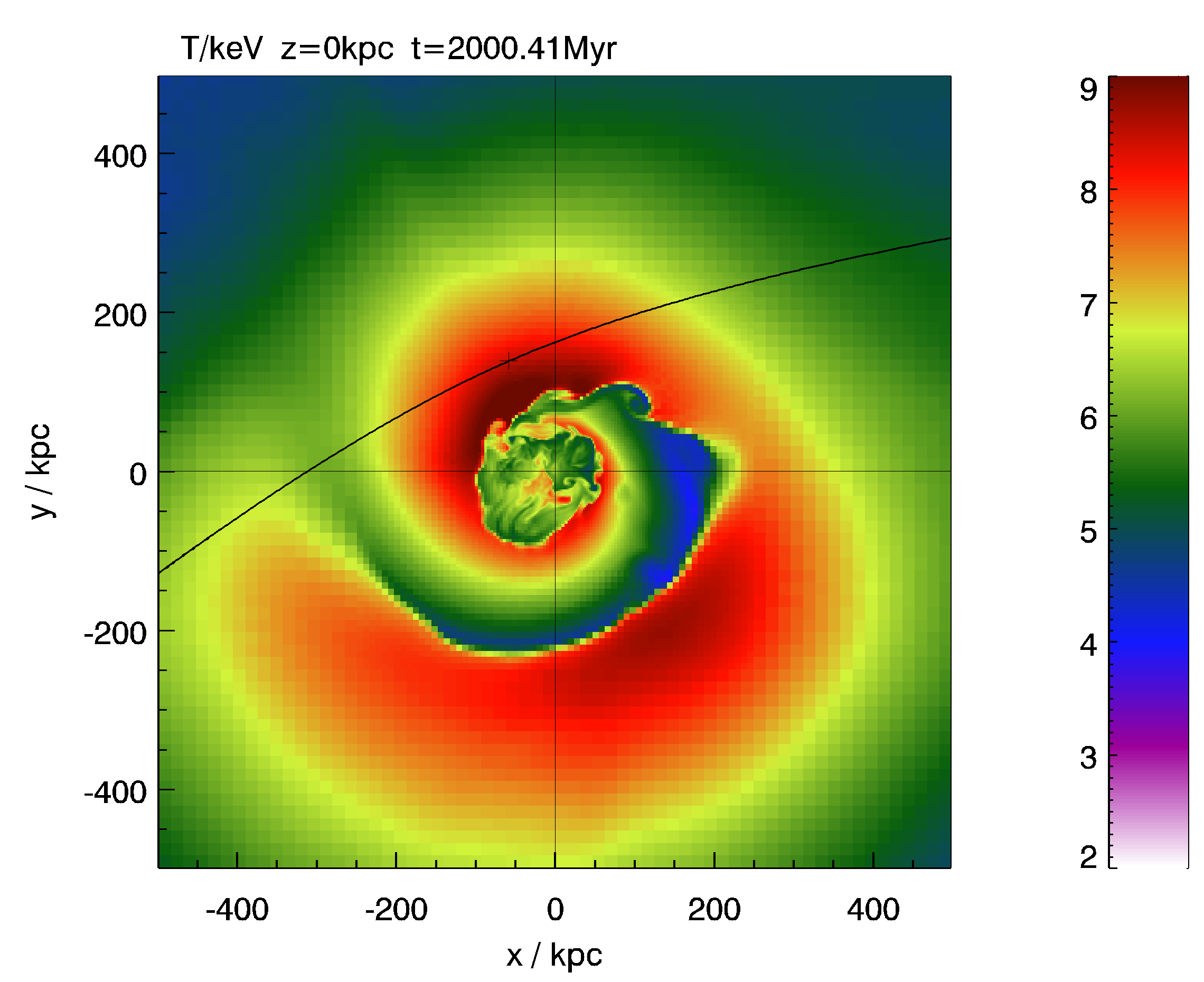}
\newline
\caption{Comparison of  hydro+Nbody simulations (first and third row) and rigid potential simulations (second and fourth row) for the merger with mass ratio 5. The panels show the ICM temperature in the orbital plane (see colour scale). The panel size is 1 Mpc. The timestep is noted above each panel. The  hydro+Nbody results are from Z10, replotted in our colour scale. The rigid potential simulation shown here uses the damping setting $(R\damp/\Kpc,L\damp/\Kpc)=(800, 300)$. The upper six panels show the onset of sloshing, the bottom six the evolution of the cold fronts and cold spiral structure. 
Our  analysis shows that after the onset  the rigid potential simulations  lag behind the hydro+Nbody one by 250 Myr (see Sect.~\ref{sec:rcf}). Hence, in the bottom six panels we plot the results from the hydro+Nbody simulations with a delay of 250\Myr.}
\label{fig:slicetemp_500kpc}
\end{figure*}

%
With both simulation methods, the gas sloshing evolves in a very similar manner. We demonstrate this in Fig.~\ref{fig:slicetemp_500kpc} by showing snapshots of the ICM temperature  in the orbital plane. The first and third row are for the hydro+N-body run, the remaining two rows for our fiducial RP simulation. 

The top two rows focus on the onset of sloshing. We see the subcluster pass the cluster centre (timesteps 0 and $1\Gyr$) from the NW over NE towards the SE. At 0.5 Gyr, sloshing has just set in, and an arc-like CF towards the S is accompanied by a cool fan towards N. Both are surrounded by hotter ICM. These properties are alike in both methods.  In the RPA, the cool fan takes a more spiral-like appearance compared to the hydro+N-body case. Also the temperature distribution S of the southern CF differs between both methods. 

The bottom six panels of Fig.~\ref{fig:slicetemp_500kpc} display the further evolution of the gas sloshing. As mentioned above, we find that  the RP simulations lag behind the hydro+N-body ones by 250 Myr for this mass ratio. Further details of this lag will be discussed in Sect.~\ref{sec:rcf}. Hence, here we plot the results from the hydro+N-body simulations with a delay of 250 Myr in order to compare corresponding timesteps.

\begin{figure}
\centering
\includegraphics[trim=0   0 0     0,clip,height=0.2\textheight]{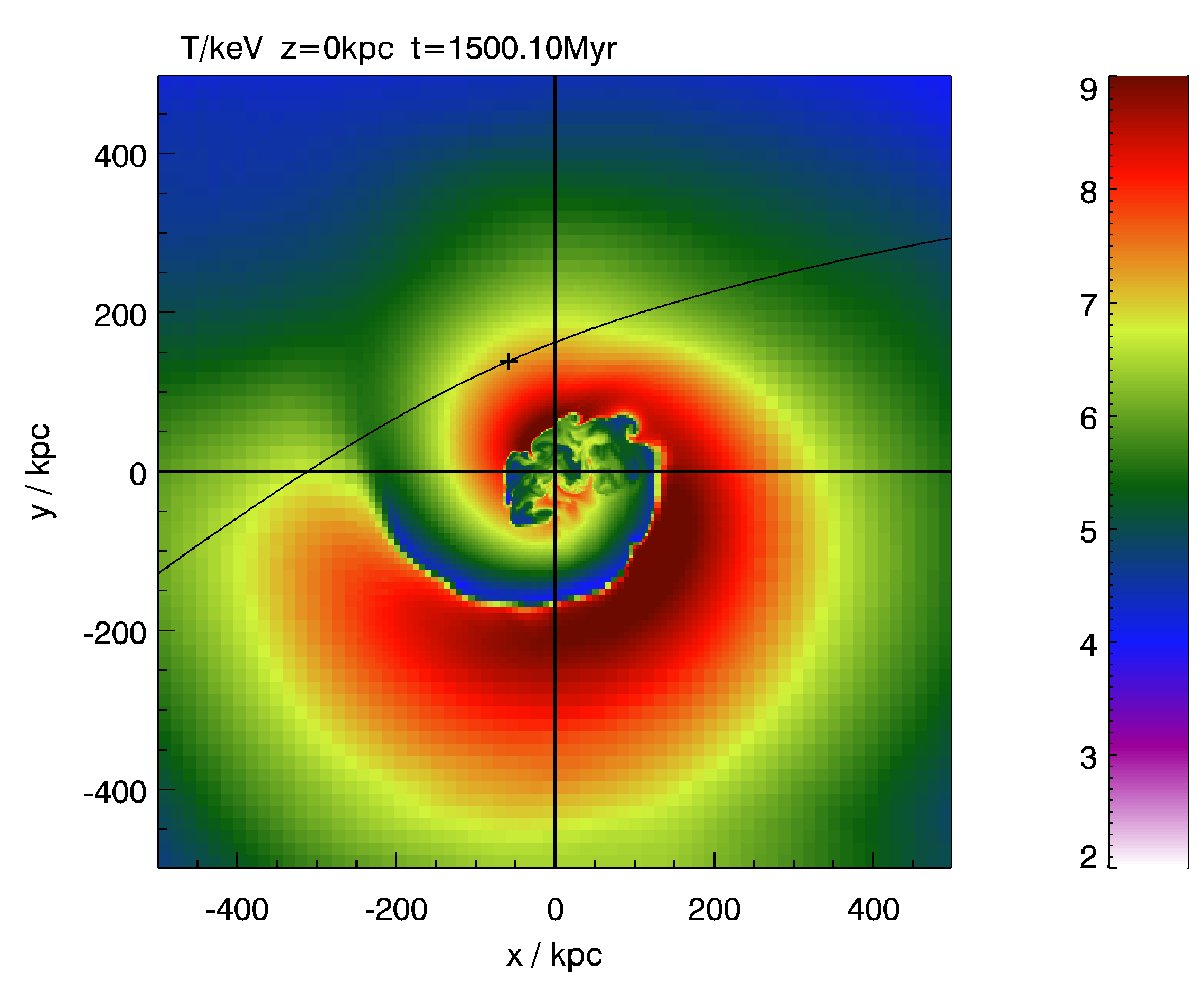}
\includegraphics[trim=0   0 0   0,clip,height=0.2\textheight]{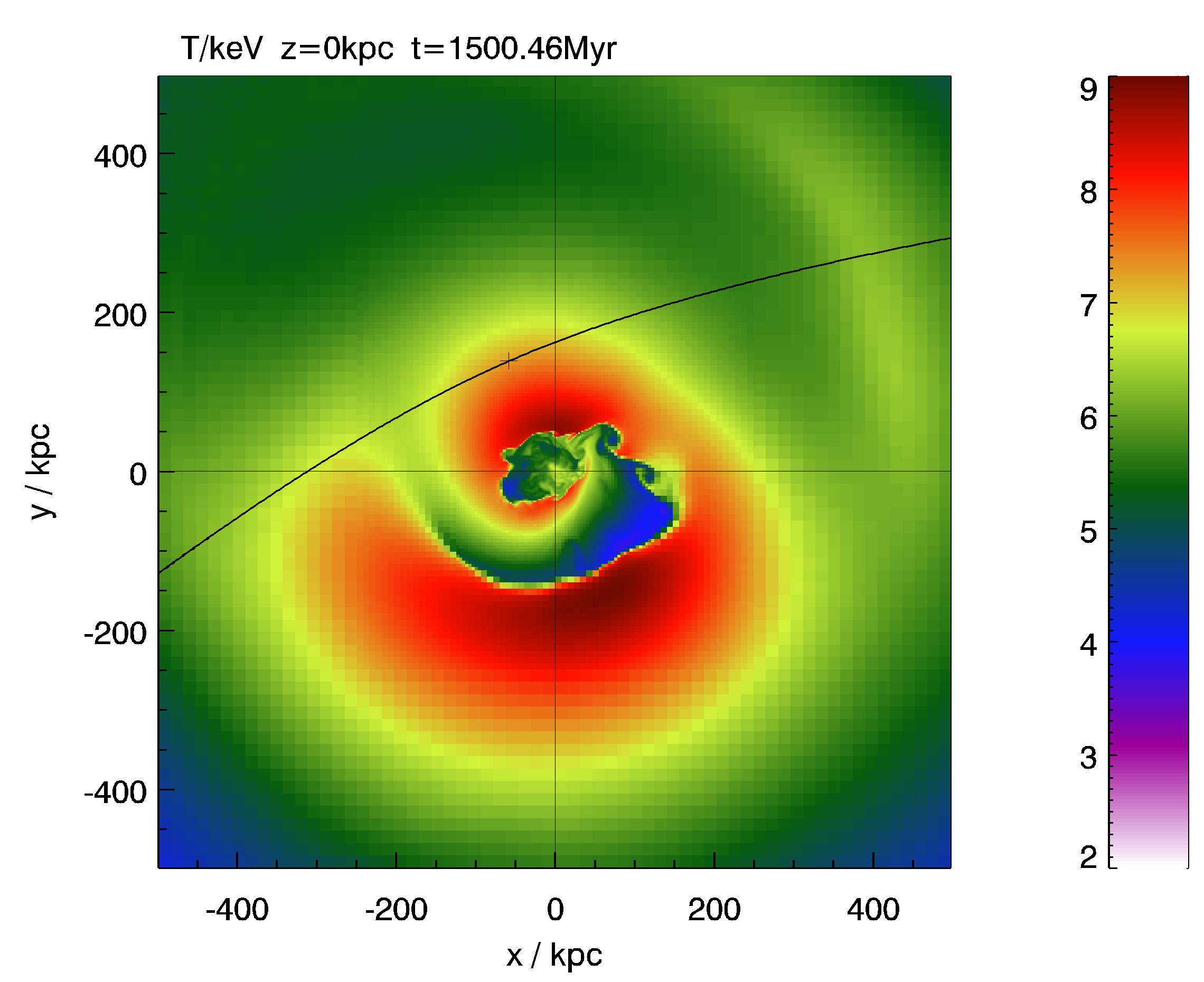}
\caption{Temperature slices for the fiducial run at $t=1.5\Gyr$, for rigid potential simulations without damping (upper panel) and with strong damping (lower panel). Compare to Fig.~\ref{fig:slicetemp_500kpc} for moderate damping and hydro+N-body case.}
\label{fig:slices_damp}
\end{figure}

In the intermediate phase (0.7 to 1.5 Gyr) the cool spiral typical for sloshing forms. The major CF is found in the SW, and a secondary CF evolves towards the NE. At its outside, the cool spiral is surrounded by a hot horse-shoe shaped region, which tends to be slightly too hot in the RPA. This evolution is the same in all cases, and the results of both methods agree well  in morphology and orientation of the cool spiral. In the late phase (2 Gyr), the CF in the SW starts to break apart in the hydro+N-body simulation. In the RP simulation, the CFs remain intact and the morphology remains close to spiral-like. We note that this break-up of the spiral structure can be recovered by using stronger damping (Fig.~\ref{fig:slices_damp}; see Sect.~\ref{sec:fiducial_damping} for a more detailed discussion of the effect of damping).

\subsubsection{Size of the cold front structure}  
The outwards motion of the CFs and hence growth of the cool spiral are important characteristics of the sloshing process. In their studies of the Virgo and Abell 496 clusters, R11a,b found that the velocity of the outward motion is largely independent of the subcluster, but is characteristic for the potential of the main cluster. This means that the  positions of the CFs in a given cluster depend mostly on the time since the subcluster's pericentre passage, i.e.~the age of the CFs. Therefore it is important to know to what extent the RPA recovers this outward motion. This is the aim of this subsection. 

\paragraph{Deriving cold front radii}  \label{sec:derive_profs}
We study the radii of the CFs towards the diagonal directions in the orbital plane. For this purpose, we first derive radial temperature profiles towards NW, SW, SE and NE, where each profile is averaged over an azimuthal extent of $\pm 15\degree$ (see Fig.~\ref{fig:compare_evolprofs_shift} for profiles and Sect.~\ref{sec:profiles} for their discussion). Given that CFs rarely form perfect circular concentric arcs around the cluster centre, this azimuthal averaging introduces a smearing out of the intrinsically discontinuous CFs over a finite radial range. This occurs even for the small azimuthal range we use for averaging.  Consequently, in the profiles, the CFs do not appear as a true discontinuity, but  as steep slopes in temperature that  stretch over typically 10 to $50\Kpc$. In each temperature profile, we  identify  CFs as regions of temperature slopes above $0.02\KeV / \Kpc$. Thus, we identify an inner and outer edge for each CF, and the nominal CF radius is defined as the average radius between this inner and outer edge. 

The first time step at which a CF can be detected is not immediately after the subcluster's pericentre passage, but typically 0.4 Gyr afterwards.  In the hydro+N-body simulation the CFs in the NW and NE direction are established only at $t=1 \Gyr$. At even later times, in all but the NE direction a second  CF at smaller radii is detectable.

\paragraph{Comparing the evolution of cold front radii} 
 \label{sec:rcf}
%
%
\begin{figure*}
\includegraphics[angle=0,width=0.7\textwidth]{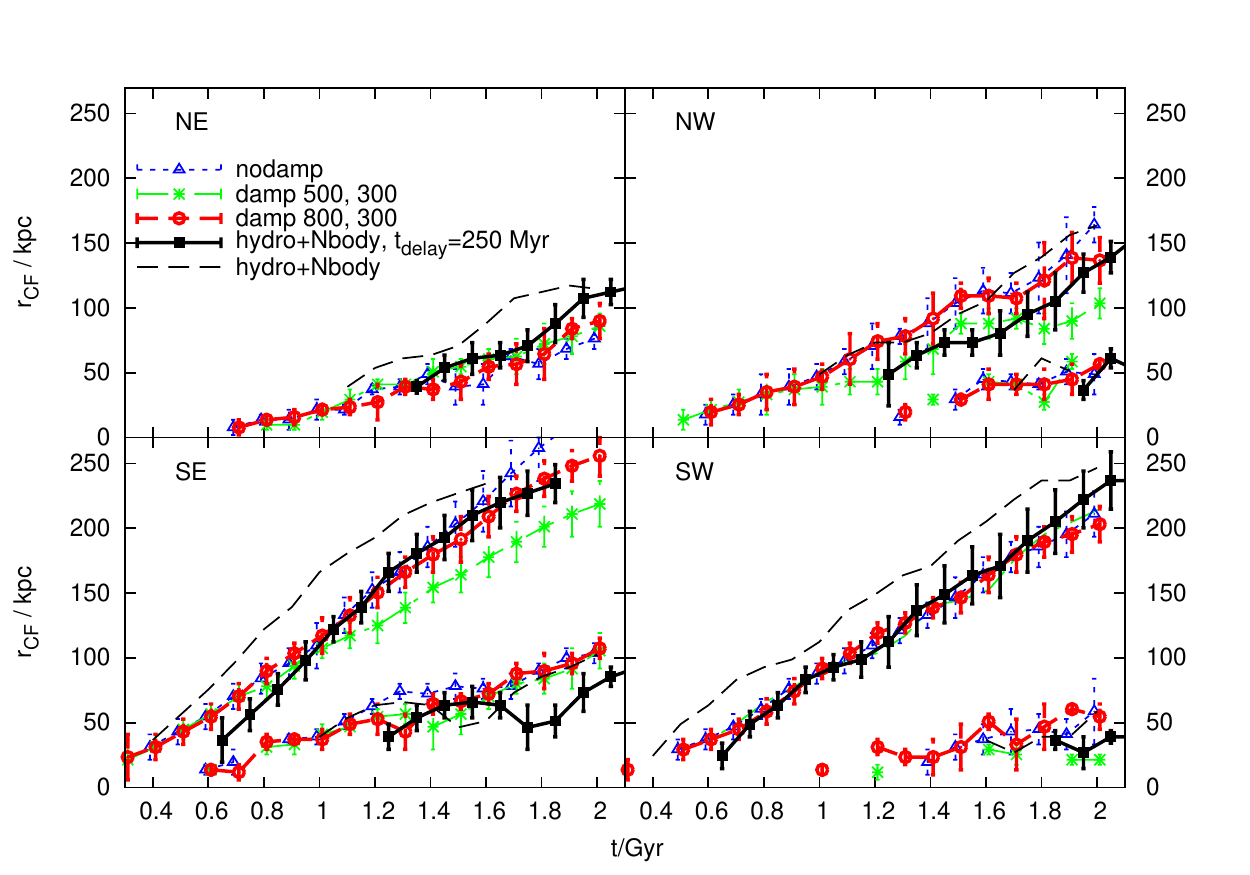}
\caption{Outward motion of the cold fronts towards the diagonal directions in the $xy$-plane (one direction per panel, see label). For the derivation of the cold front radii, see Paragraph~\ref{sec:derive_profs}. The error bars indicate the width of the cold front as it appears in the azimuthally averaged profiles. In all but the NE panel, the evolution of the outermost and second cold front is shown. The figure compares the results for different realisations of the mass ratio 5 merger. The black dashed line without error bars represents the hydro+N-body run, the other coloured/broken lines are for the results from the rigid potential simulations with different damping settings (see legend). Clearly, the rigid potential simulations lag behind the hydro+N-body one. Plotting this reference with a delay of $250\Myr$ (and remembering footnote 1) corrects for this lag.}
\label{fig:evol_CFradii}
\end{figure*}
%

Having   derived the positions of the CFs at each timestep, we can now proceed to analyse their outward motion. We do so in Fig.~\ref{fig:evol_CFradii}, where we plot the temporal evolution of the CF radii towards the diagonal directions (NE, NW, SE, and SW) in the orbital plane. We use error bars to indicate the width of each CF, i.e.~from its inner to outer edge. The RP simulations with different settings are plotted by coloured lines. Here we focus on the red line, which marks our fiducial RP run. 

The result from the hydro+N-body simulation is the dashed black line. The RP results differ systematically from this reference: At a given timestep, the RP simulations produce somewhat too small radii for the outermost CFs. This is true for all directions except NW. Thus, in general, the RP simulations lag behind the hydro+N-body one. Plotting the hydro+N-body result with a delay of 250 Myr (solid black line with error bars, and remember footnote 1) compensates the difference in all but the NW direction, leading to a good agreement to the RP simulation. The second CFs agree well between all runs.

This lag in CF motion is the major systematic difference between two simulation methods. Consequently, we use the delay of 250 Myr for the full hydro+N-body simulation in all other comparisons  regarding the fiducial merger case.

\subsubsection{Quantitative comparison of density and temperature distribution} 
\label{sec:profiles}
\begin{figure*}
\includegraphics[angle=0,width=\textwidth]{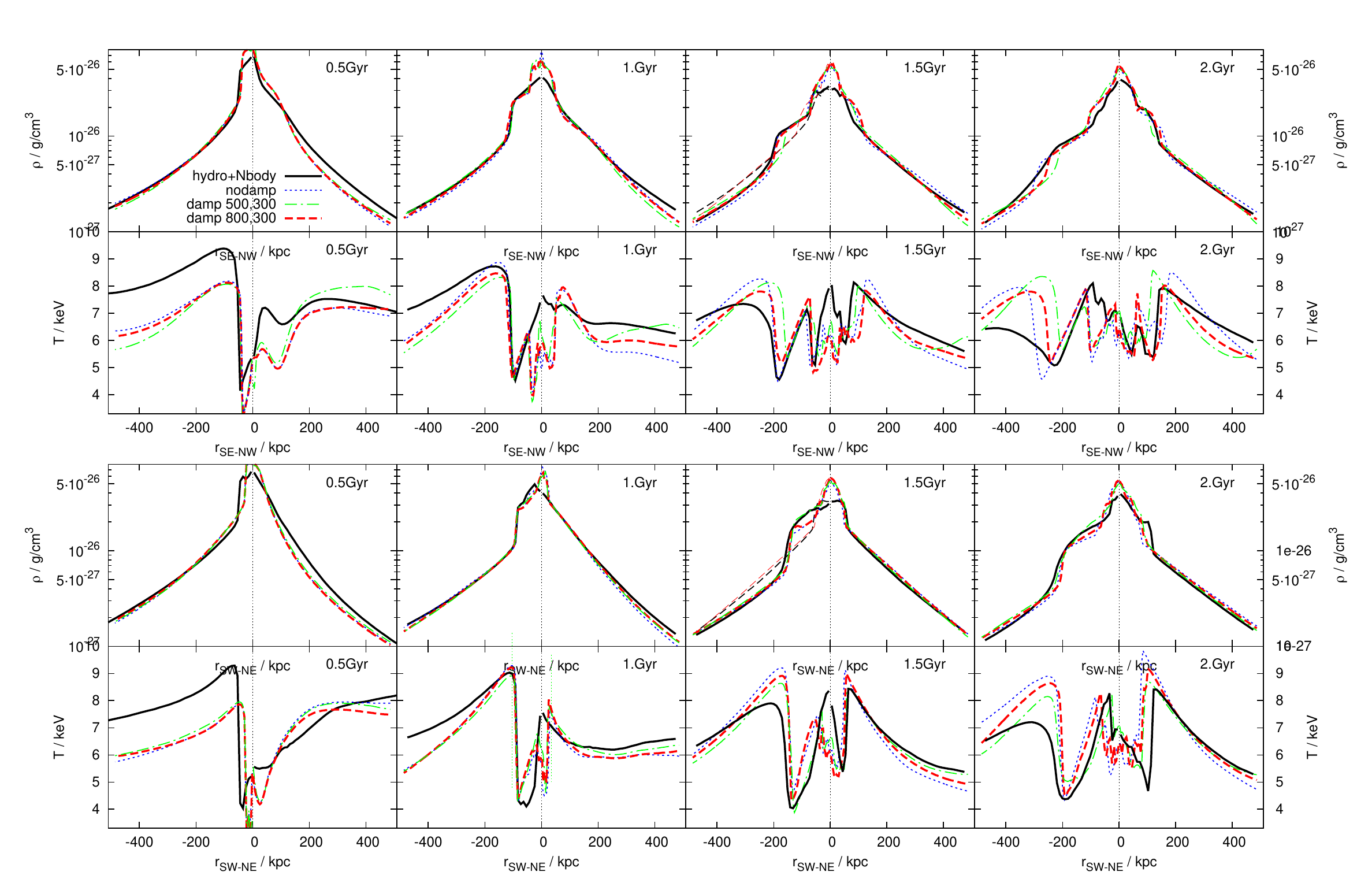}
\caption{
Comparison of density and temperature profiles along diagonals in the $xy$-plane. The columns are for different timesteps as indicated in each panel. The two top  rows are for the SE-NW direction, the two bottom  rows for the SW-NE direction. The black solid lines show the hydro+N-body simulation, coloured broken lines lines the rigid potential simulations with different damping.  Profiles are averaged over $\pm 15\degree$ around the indicated direction.  The hydro+N-body simulation is plotted with a delay of $250\Myr$ in all but the $0.5$ Gyr step, which accounts for the lag of rigid potential approximation described in Sect.~\ref{sec:rcf}. For the density profiles at $t=1.5 \Gyr$, we overplot the profile of the positive direction as thin lines in the negative direction in order to demonstrate the asymmetry.}
\label{fig:compare_evolprofs_shift}
\end{figure*}

In order to go beyond the qualitative comparison of the temperature slices in Fig.~\ref{fig:slicetemp_500kpc}, we derive radial temperature and density profiles as described above in Paragraph~\ref{sec:derive_profs}. In  Fig.~\ref{fig:compare_evolprofs_shift} we compare these profiles for different realisations for the fiducial merger. 
%
%
Again, here we concentrate on the red and black lines, which are for the fiducial RP simulation and the hydro+N-body one, respectively. 

At 0.5 Gyr, the sloshing is still in the onset phase and we do not apply the delay discussed above. Still, already here the CF in the SW is ahead in the hydro+N-body simulation.  Here, the RPA run shows a weaker impact of the subcluster passage on the temperature. Also the density profile in the northern directions are not accurately reproduced. This is the region the subcluster directly passes and the strongest differences are to be expected. 

In all later timesteps we apply the delay of 250 Myr  to the hydro+N-body simulation as derived in  Sect.~\ref{sec:rcf}. As a result, we  achieve a good agreement between both methods. Especially along the SW-NE axis, which is perpendicular to the subcluster orbit, the agreement is excellent. The only systematic difference between both methods is that the RPA over-predicts  the temperature just outside the CFs, which we have already seen in Fig.~\ref{fig:slicetemp_500kpc}.

\subsubsection{Density and temperature across the CFs}
%
\begin{figure*}
\includegraphics[angle=0,width=0.7\textwidth]{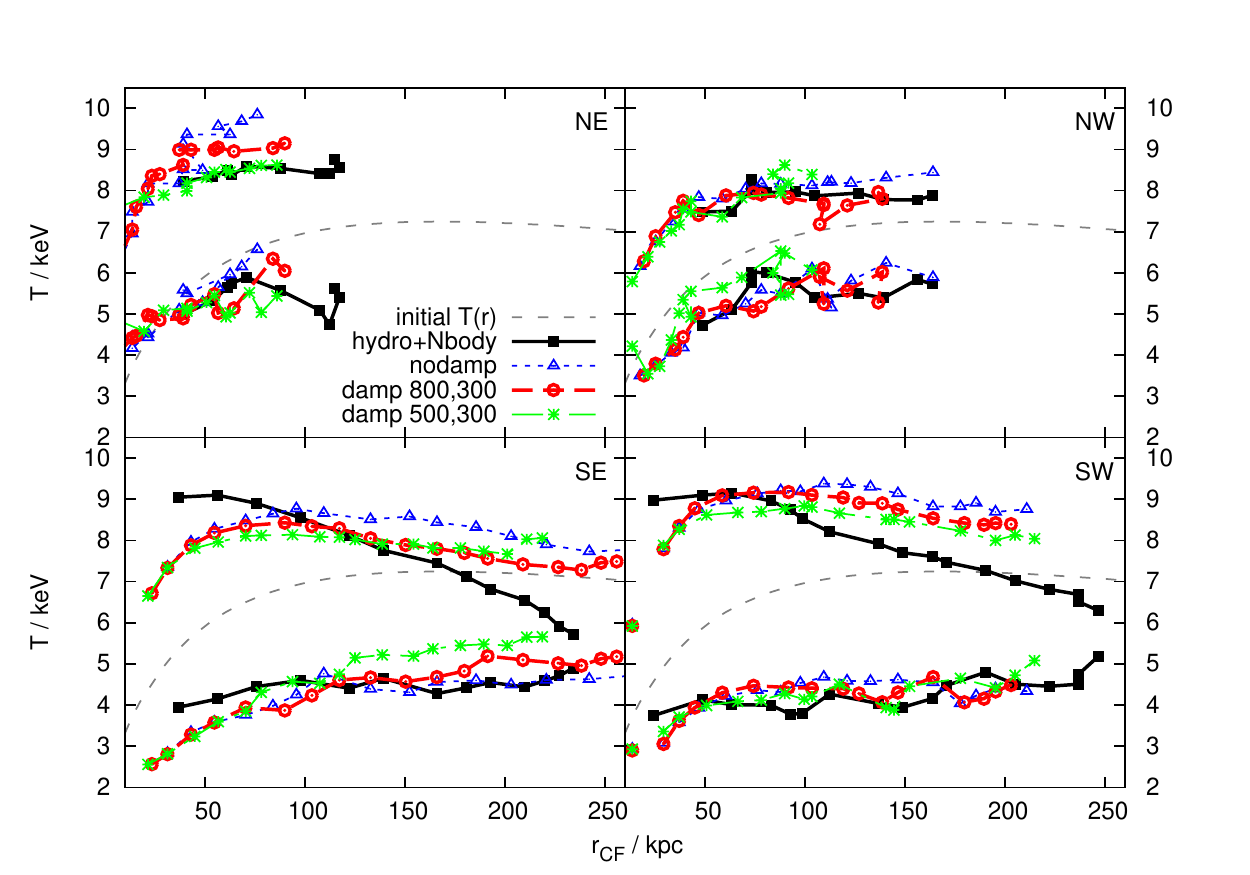}
\caption{Temperature inside and outside of the cold front as a function of cold front radius. We show results for the four diagonal directions in the $xy$-plane. Different line colours/styles code different realisations of the mass ratio 5 merger, see legend. In each panel, the upper and lower sets of lines denote the temperatures just outside and inside the cold fronts, respectively. For comparison, we plot the initial temperature profile as the dotted line.}
\label{fig:evol_CFradii_Temp}
\end{figure*}
%
\begin{figure*}
\includegraphics[angle=0,width=0.7\textwidth]{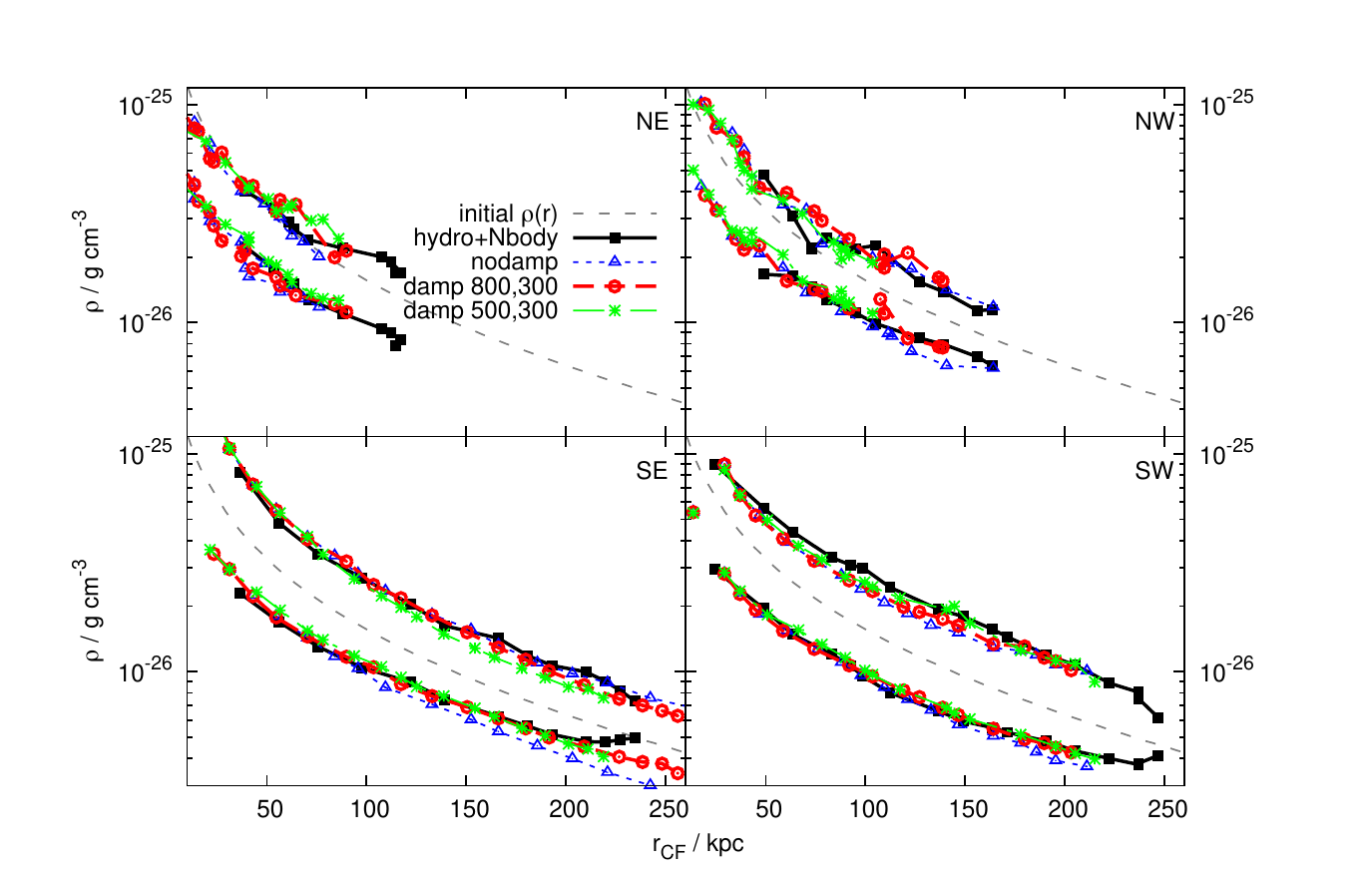}
\caption{Same as Fig.~\ref{fig:evol_CFradii_Temp}, but for density inside  and outside of the cold fronts (upper and lower set of lines, respectively).}
\label{fig:evol_CFradii_dens}
\end{figure*}

Here we aim at  comparing the evolution of the temperature and density at the CFs. For this purpose, we derive both quantities at the inner and outer edge of each CF along with its radius (see Paragraph~\ref{sec:derive_profs}).  We note that the combination of azimuthal and radial binning introduces an uncertainty in temperature of at least $\pm0.5$ keV at both edges.  In Fig.~\ref{fig:evol_CFradii_Temp} we plot the temperature at the inner and outer edge of each CF as a function of its radius. Thus, we compare simulations at stages when the CF spiral has reached the same size. We compare the results along the diagonal directions in the orbital plane.  This figure is restricted to the outermost ring of CFs. Also here, we focus on the red and black lines, which are for the fiducial RP simulation and the hydro+N-body one, respectively. 

The temperatures at the inner edges of the CFs are reproduced well.  The outer temperatures in the SE and SW directions differ systematically between the rigid potential and the hydro+N-body simulation. In these positions, the rigid potential simulations first underestimate the outer temperature and  overestimate it at later times.  The reason for this difference is that the rigid potential approximation leads to a different flow field outside the central region of the cluster. In all but the early times, this leads to a stronger compressional heating at the outer side of the CFs and thus a higher temperature. This systematic difference can be lessened somewhat by the damping, but is not prevented completely. 

We apply the same analysis to the density inside and outside each CF and present the result in Fig.~\ref{fig:evol_CFradii_dens}.  For the fiducial RP simulation, the densities at the CFs are reproduced well. Thus, even at late stages, where the RPA does not accurately estimate the temperatures outside the CFs, it is still accurate for the  densities.

\subsubsection{Large-scale asymmetry}
In addition to the cold spiral structure in the cluster centre, also the large-scale distribution of the ICM density at least out to 500 kpc is reproduced well in the RPA. Both simulation methods find the characteristic asymmetry  in the sense that profiles of, both, density and temperature on opposite sides of the cluster centre alternate around each other, switching over at the CFs. We have illustrated this effect in Fig.~\ref{fig:compare_evolprofs_shift} in the density panels for $t=1.5 \Gyr$ by overplotting the profiles from the NE side to the SW direction and the NW ones to the SE direction. 

Qualitatively, the large-scale structure in temperature is reproduced, too. However, here the details depend on the damping (see Sect.~\ref{sec:fiducial_damping}).

\subsection{Impact of damping} \label{sec:fiducial_damping}
%
%
%
The damping  of the inertial frame correction described in Sect.~\ref{sec:method_damping} is constructed such that it gradually switches off the inertial frame correction in the cluster outskirts, which should not be applied there. Using the RPA without this "re-correction" leads to an unrealistic ICM velocity field in the outer cluster regions and two artefacts compared to the  hydro+N-body reference run:  a cooler temperature in the outer northern region and the hotter temperature outside all CFs. Both effects can be seen in the temperature slices (top panel of Fig.~\ref{fig:slicetemp_500kpc}) and all other comparison plots (Figs.~\ref{fig:compare_evolprofs_shift}, \ref{fig:evol_CFradii_Temp} and \ref{fig:evol_CFradii_dens}). 

Using a strong damping  of $(R\damp/\Kpc,L\damp/\Kpc)=(500, 300)$ leads to a nearly correct temperature distribution in the outer cluster regions as demonstrated in the temperature slice in Fig.~\ref{fig:slicetemp_500kpc}. However, the temperature just outside the CFs is still slightly too high (Figs.~\ref{fig:slicetemp_500kpc}, \ref{fig:compare_evolprofs_shift}, \ref{fig:evol_CFradii_Temp}). Moreover, the strong damping leads to slightly smaller CF radii towards the  NW and SE direction after $t=1\Gyr$,  which is the direction of motion of the subcluster (Fig.~\ref{fig:evol_CFradii}). The  CF radii towards the NE and SW, i.e.~along the axis perpendicular to the orbit,  are independent of damping  at all times. 

In summary, the stronger the damping, the more accurate is the outer temperature distribution, but at some expense of the accuracy in the cluster centre. Hence, we prefer to use a moderate damping which ensures an accurate reproduction of the CF radii, temperatures inside them and densities at, both, the inside and outside. 

The case (800, 100) is very similar to our fiducial setting (800, 300) demonstrating that the results are not sensitive to the choice of the fall-off length scale for the damping, $L\damp$.

\subsection{Subcluster orbit and mass evolution} \label{sec:orbit}
%
\begin{figure}
\centering\includegraphics[angle=0,width=0.35\textwidth]{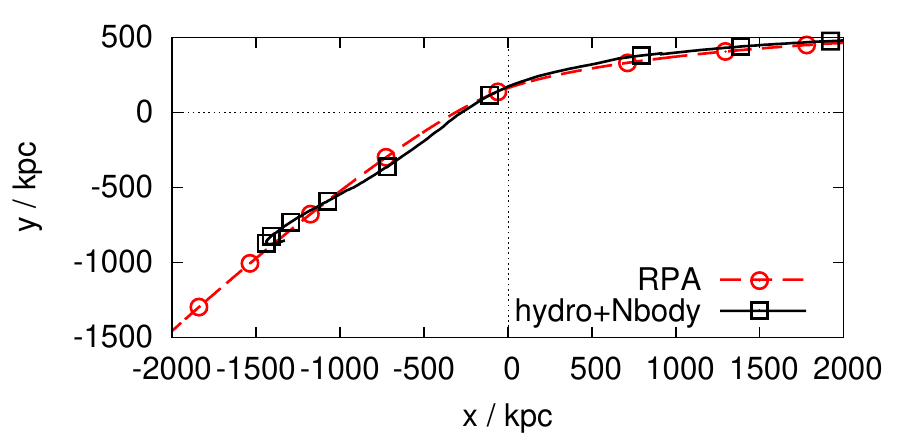}
\caption{Comparison of the subcluster orbit in the hydro+N-body simulation (black solid line) and the rigid potential approximation (red dashed line) for the merger with mass ratio 5. We mark the subcluster positions in 250 Myr intervals. While the direction of the orbit is reproduced well in the rigid potential approximation, the neglect of the dynamical friction leads to significant over-estimation of the velocity after 0.5 Gyr after pericentre passage.}
\label{fig:orbits}
\end{figure}
%
In addition to the ICM properties, we also compare  the subcluster orbit w.r.t.~the cluster centre from both methods in Fig.~\ref{fig:orbits}. The test particle orbit used in the RP simulations approximates the true trajectory of the subcluster very well. We have marked the subcluster position in steps of 250 Myr along both orbits, demonstrating that the test particle approximation also recovers the motion of the test particle along its orbit up to a few 100 Myr after pericentre passage. After that,  dynamical friction causes the subcluster to decelerate significantly, such that its new apocentre distance is only about 1.8 Mpc, and the apocentre is reached already 1.25 Gyr after pericentre passage. The test particle method does not capture this effect, and predicts a comparable cluster-centric distance already after about 0.6 Gyr after pericentre passage. \citet{Taylor2001} proposed an algorithm to incorporate dynamical friction in a simplified form.  However, the  gas sloshing we are interested in  is triggered by the pericentre passage and thus is unaffected by this difference in subcluster position at later epochs. 

\begin{figure}
\centering\includegraphics[angle=0,width=0.3\textwidth]{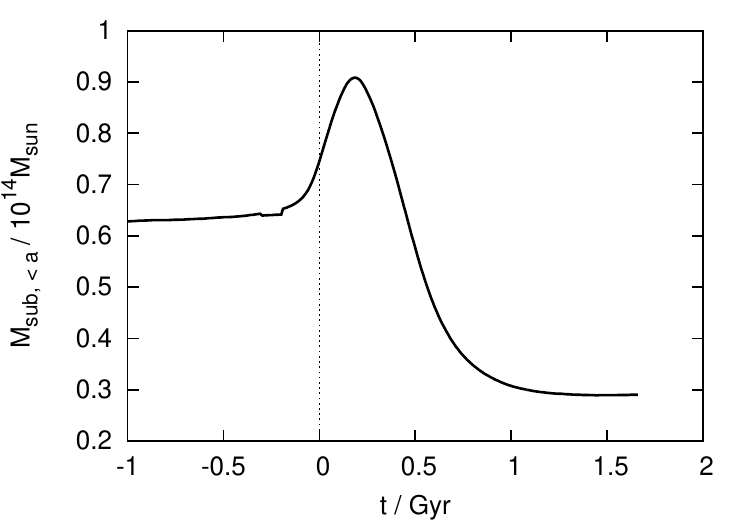}
\caption{Evolution of subcluster mass within the subcluster scale radius. During pericentre passage and up to 0.2 Gyr afterwards the subcluster suffers tidal compression. It  starts loosing mass significantly only 0.5 Gyr after pericentre passage.}
\label{fig:subclustermass}
\end{figure}

In Fig.~\ref{fig:subclustermass} we demonstrate the evolution of the subcluster by plotting its mass within its scale radius. Clearly,  during pericentre passage and up to 0.2 Gyr afterwards the subcluster suffers tidal compression. It  starts loosing mass significantly only 0.5 Gyr after pericentre passage. This is well after triggering the gas sloshing and thus has no significant effect on the further evolution of the ICM in the main cluster. We will discuss the differences in the evolution of the main cluster potential in Sect.~\ref{sec:discuss_delay}.
%
%

\subsection{Mergers with other mass ratios}
%
\subsubsection{Mass ratio 20}
%
\begin{figure}
\includegraphics[trim=0   0 450     0,clip,height=0.175\textheight]{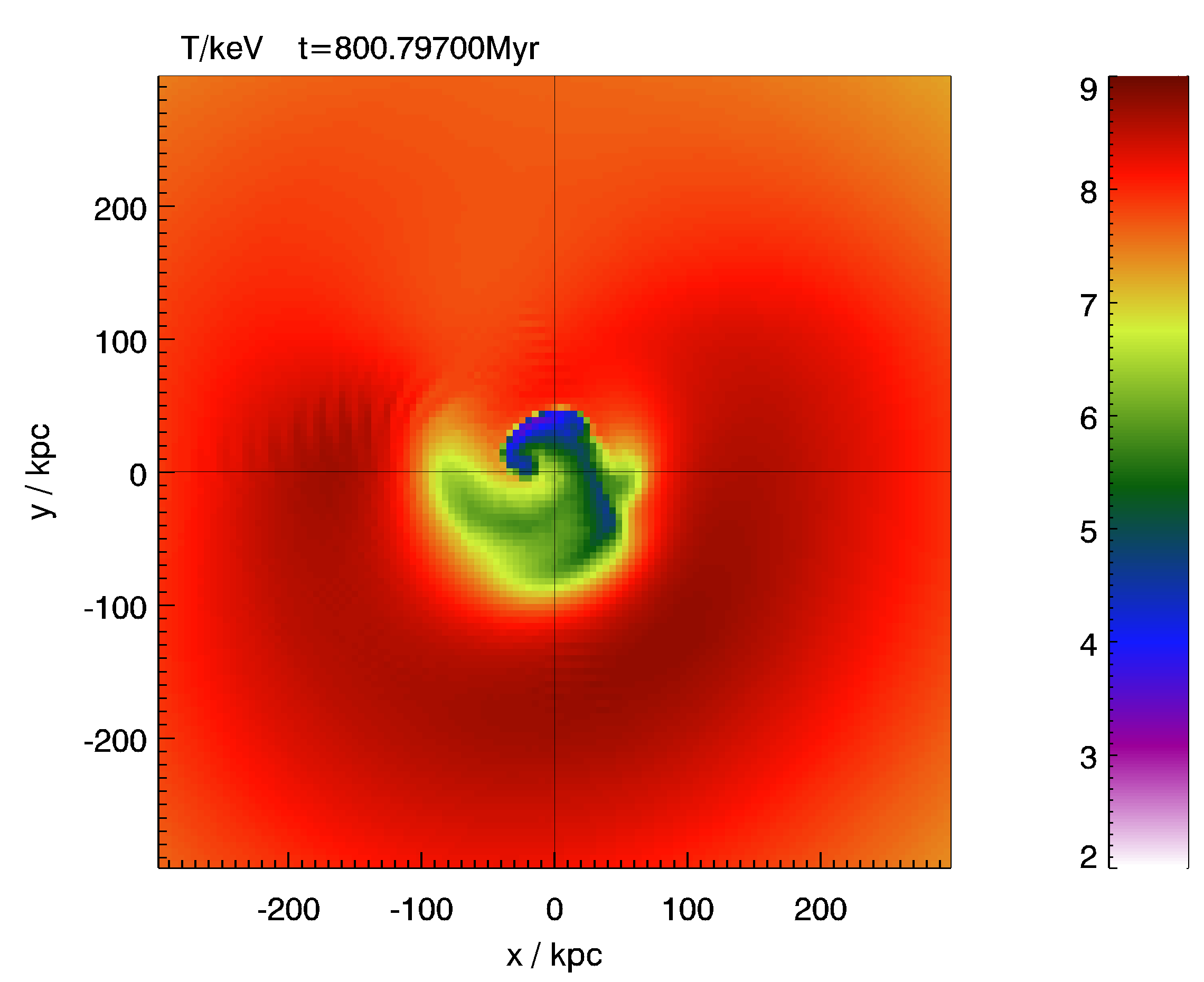}
\includegraphics[trim=300   0 450     0,clip,height=0.175\textheight]{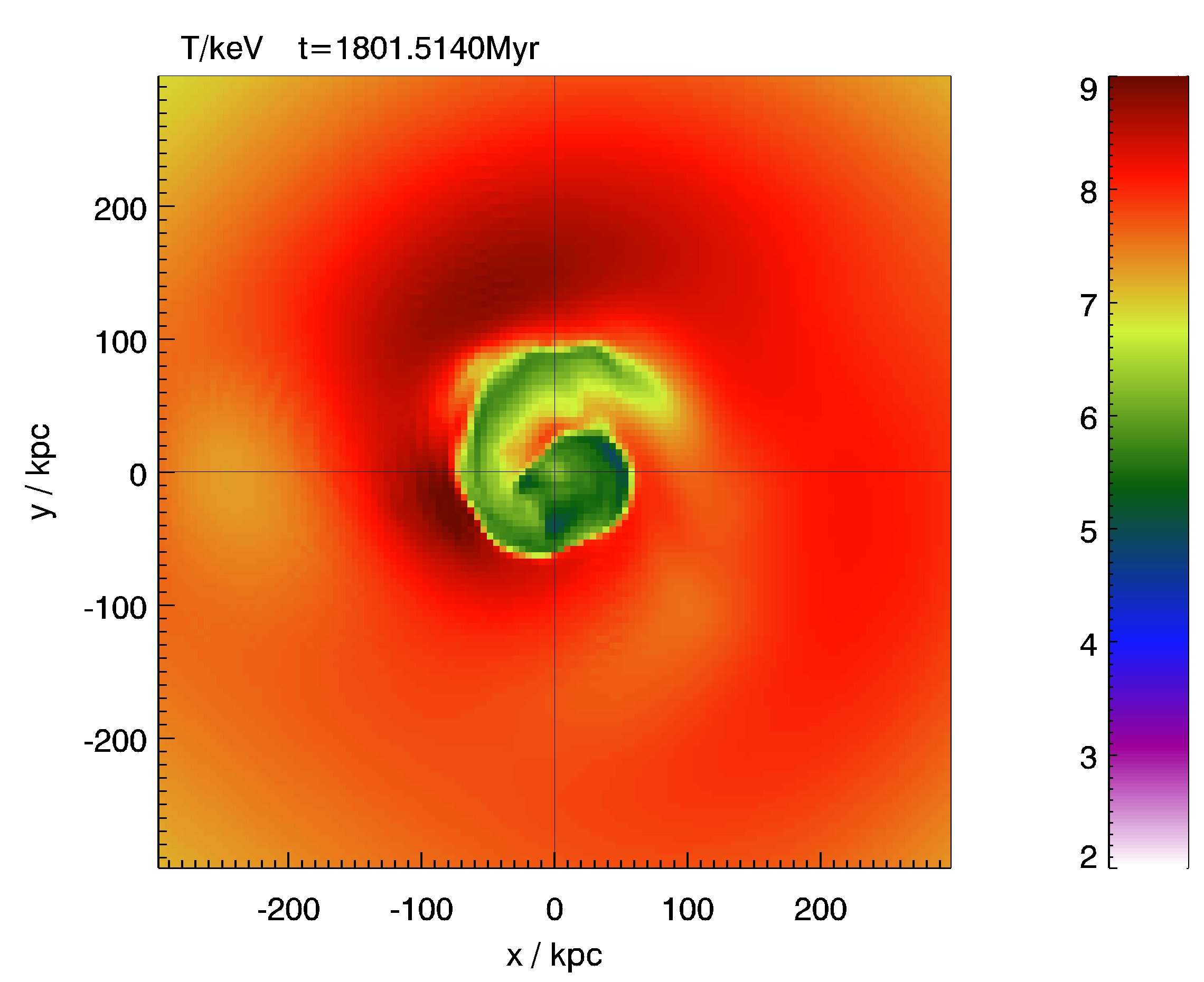}
\hspace{-0.5cm}
\includegraphics[trim=2000   0 50     0,clip,height=0.175\textheight]{PLOTS/JOHN_R20_r200/slice_Temp_z_0_0314.png}
\newline
\includegraphics[trim=0   0 450     0,clip,height=0.175\textheight]{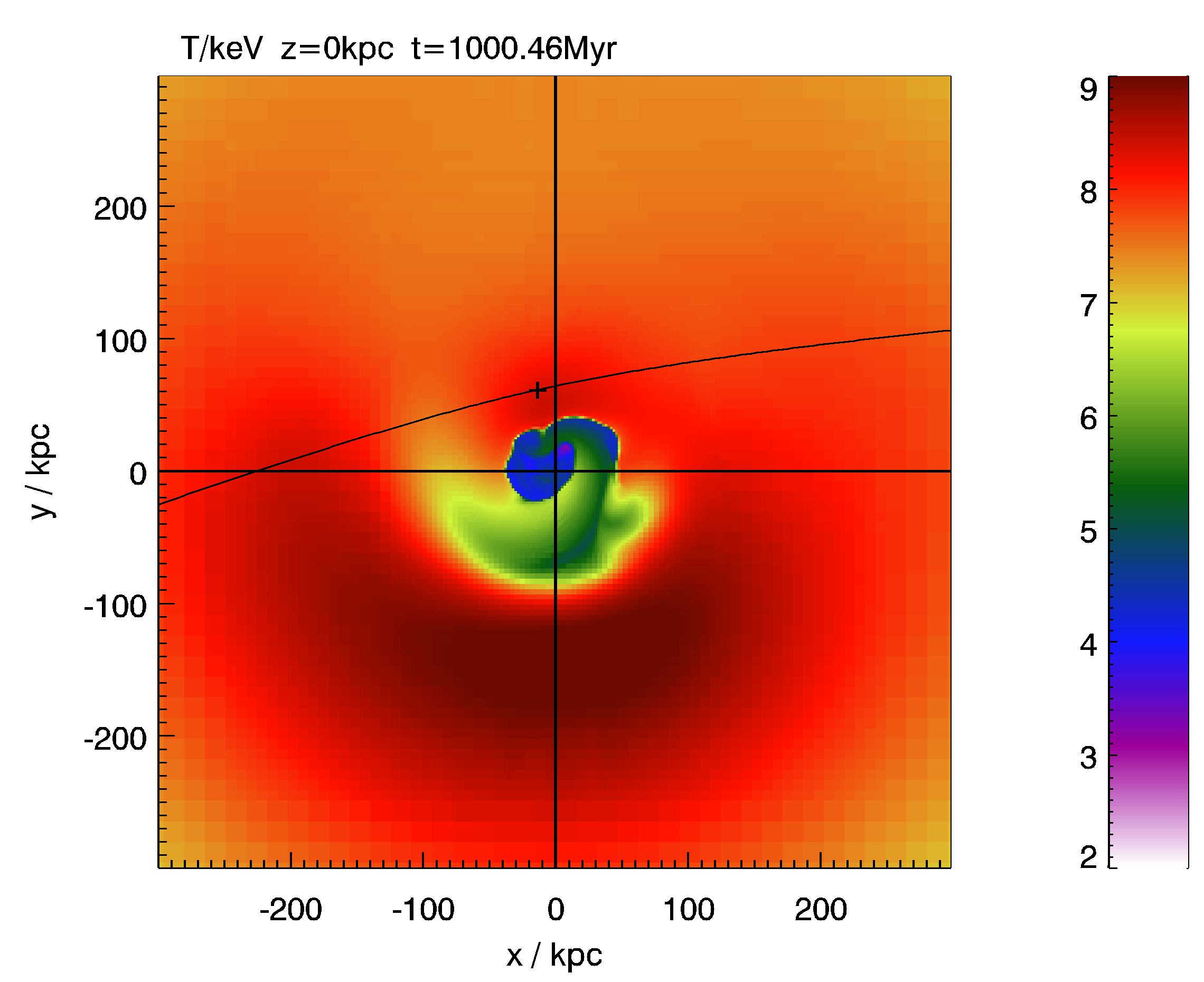}
\includegraphics[trim=300   0 450     0,clip,height=0.175\textheight]{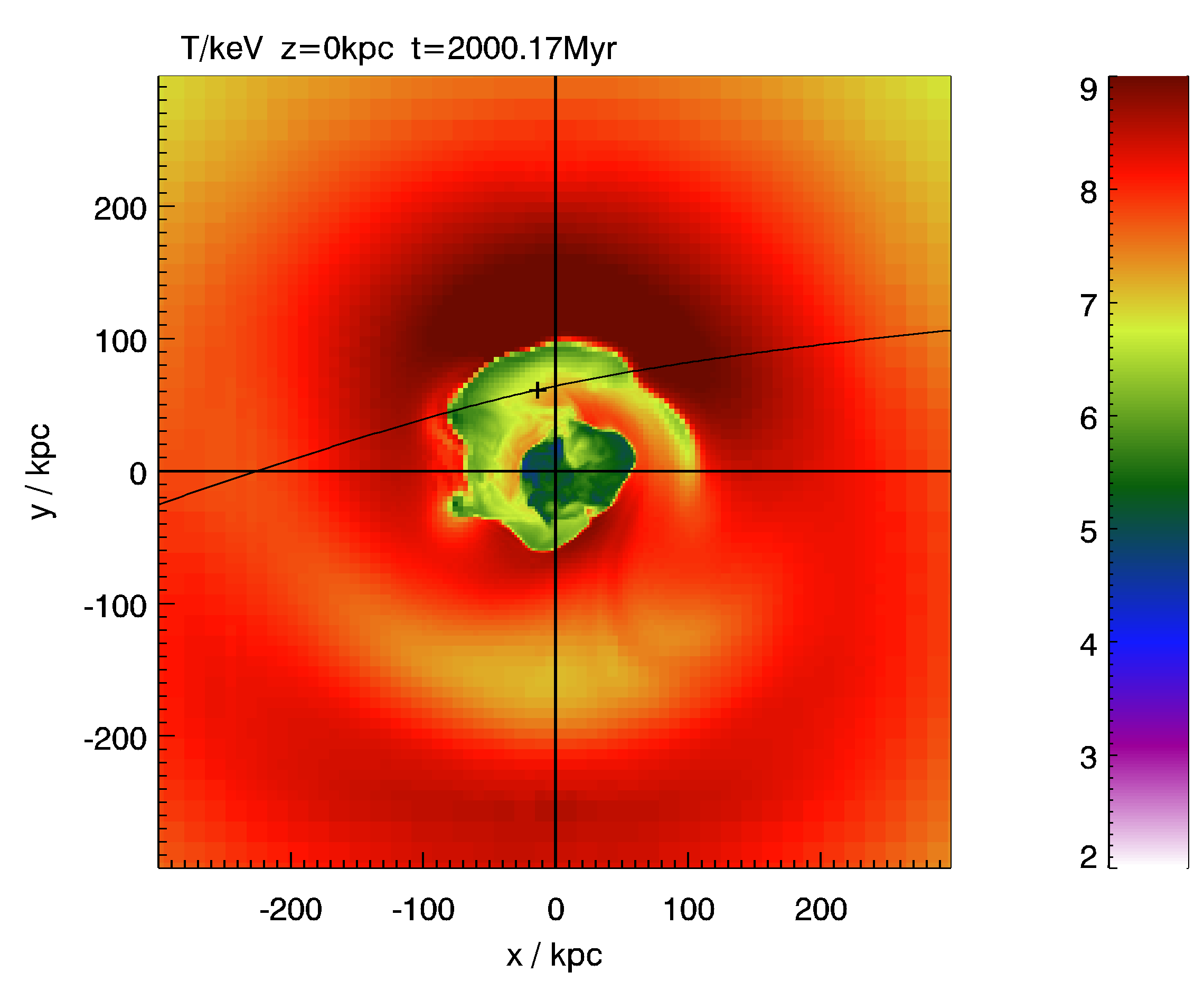}
\newline
\caption{Snapshots of the temperature in the orbital plane for the merger with mass ratio 20. We show the hydro+N-body results (upper row) with a delay of 200 Myr compared to the rigid potential run (bottom row). Both methods give very similar results.}
\label{fig:slicetemp_R20}
\end{figure}
%
\begin{figure}
\includegraphics[angle=0,width=0.5\textwidth]{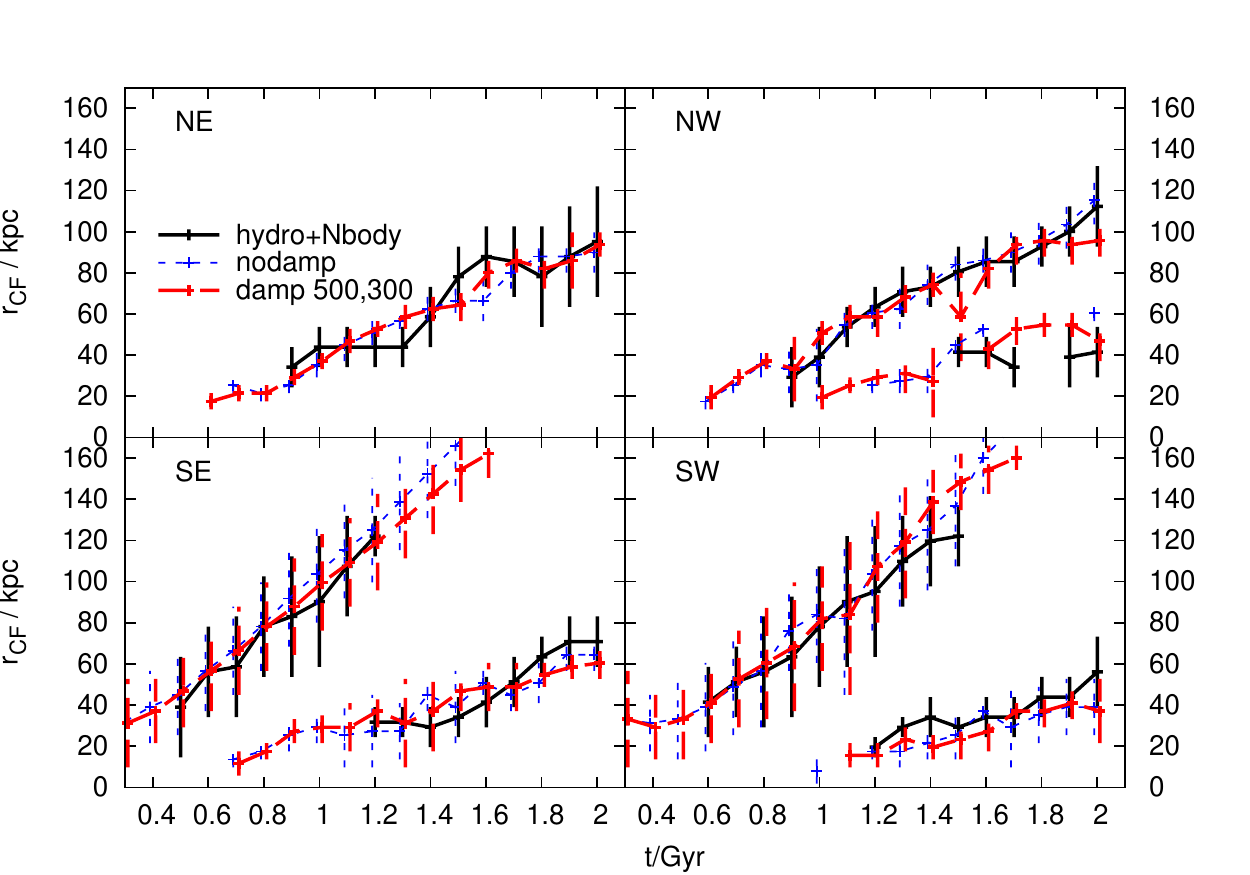}
\caption{Same as Fig.~\ref{fig:evol_CFradii}, but for the merger with a mass ratio of 20. The result from the hydro+N-body simulation is plotted with a delay of $200\Myr$.}
\label{fig:evol_CFradii_shift_R20}
\end{figure}
%
\begin{figure}
\includegraphics[angle=0,width=0.5\textwidth]{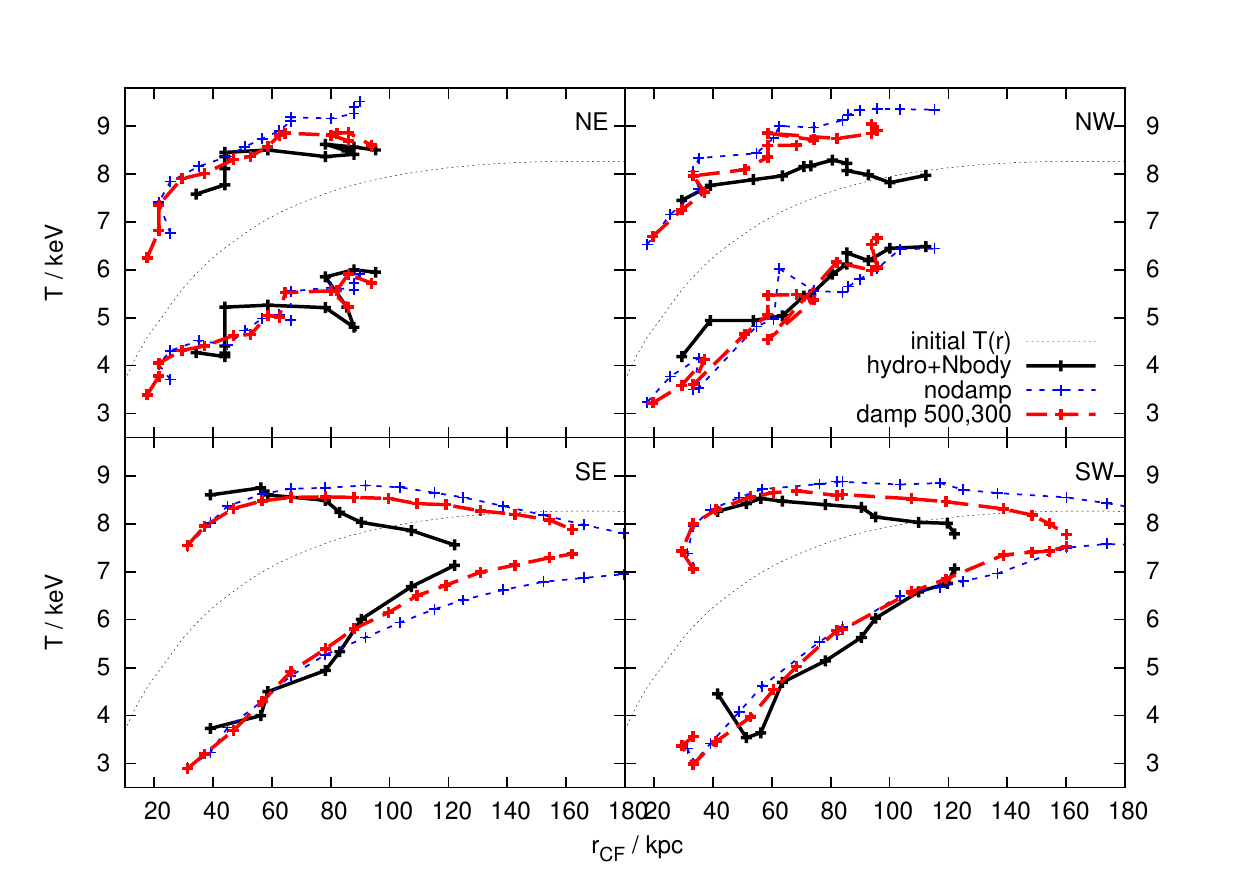}
\caption{Same as Fig.~\ref{fig:evol_CFradii_Temp}, but for the merger with a mass ratio of 20.}
\label{fig:evol_CFradii_Temp_R20}
\end{figure}

We have run the same comparison for a minor merger with a larger mass ratio of 20. Here, the subcluster has a smaller scale radius of only 220 kpc, hence the damping setting $(R\damp/\Kpc,L\damp/\Kpc)=(500, 300)$ corresponds to moderate damping here.  In Fig.~\ref{fig:slicetemp_R20} we compare temperature slices, in Fig.~\ref{fig:evol_CFradii_shift_R20} the evolution of the CF radii, and the temperatures inside and outside each CF in Fig.~\ref{fig:evol_CFradii_Temp_R20}. Even for this case with a much smaller subcluster, the sloshing in the RPA runs lags behind the hydro+N-body one by 200 Myr. Hence, we come to the same results and conclusions as in the fiducial case.

\subsubsection{Mass ratio 2}
%
\begin{figure*}
\includegraphics[trim=300   0 300     0,clip,height=0.175\textheight]{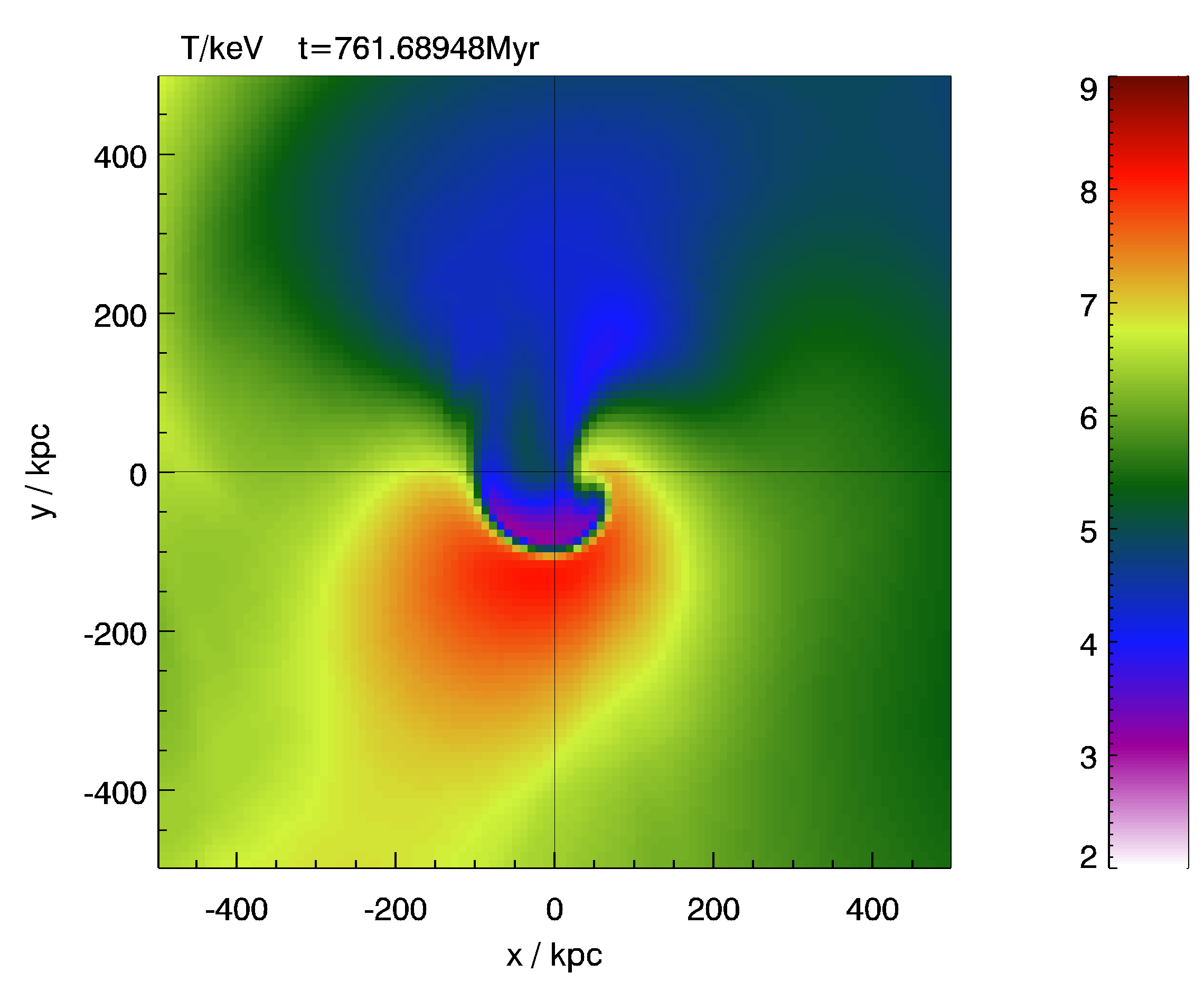}
\includegraphics[trim=300   0 300     0,clip,height=0.175\textheight]{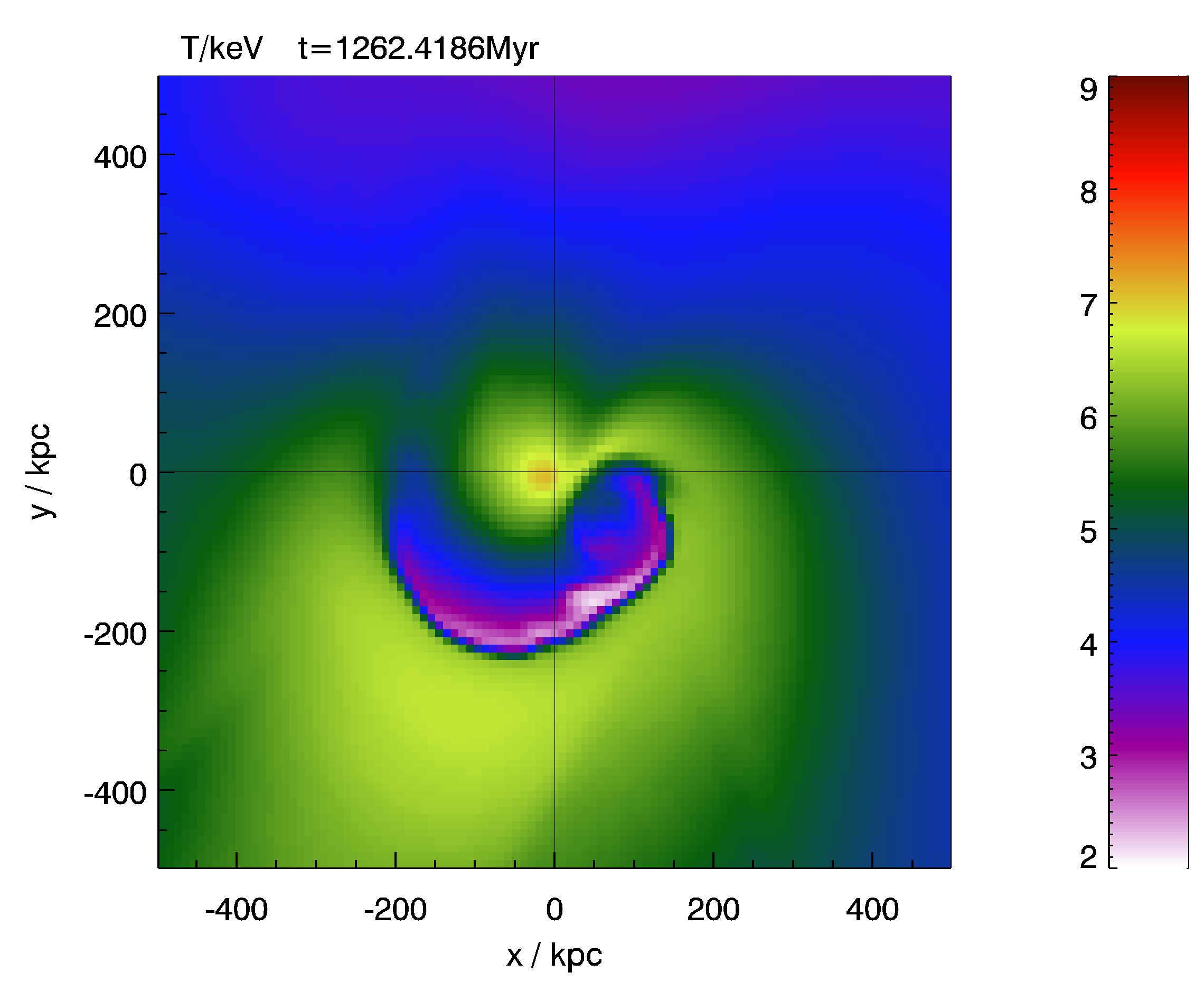}
\includegraphics[trim=300   0 0     0,clip,height=0.175\textheight]{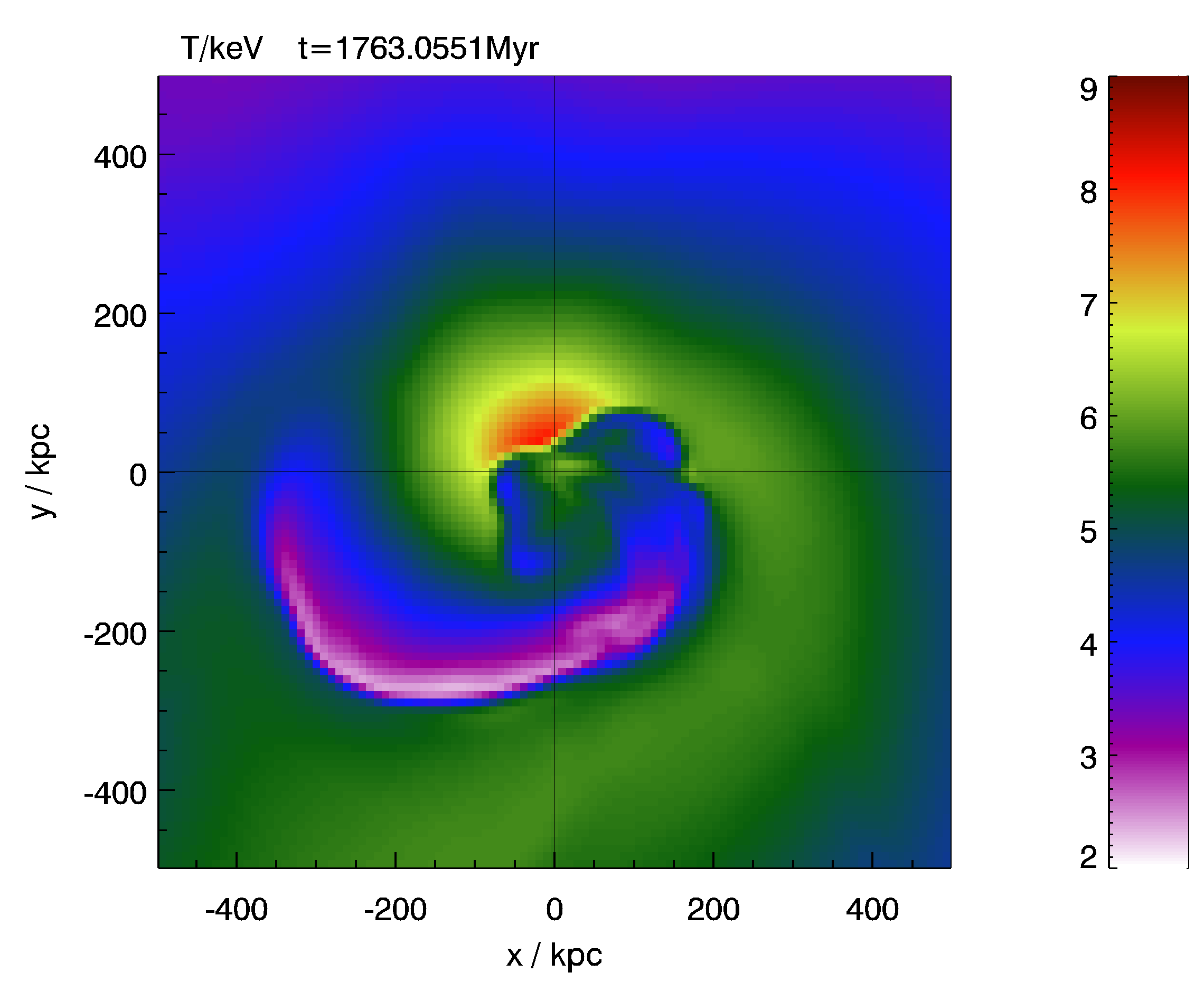}
\newline
\includegraphics[trim=300   00 300     0,clip,height=0.175\textheight]{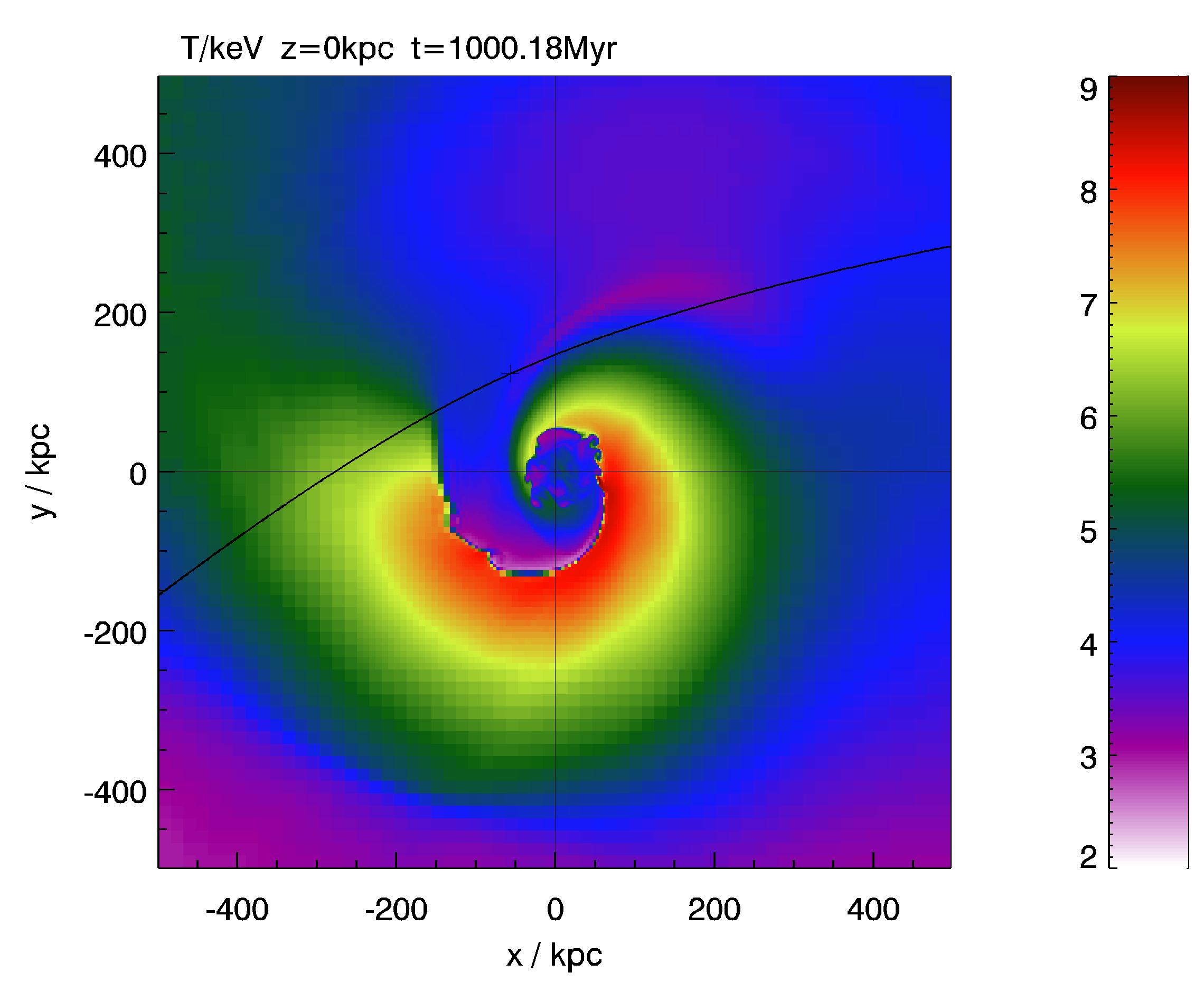}
\includegraphics[trim=300   00 300     0,clip,height=0.175\textheight]{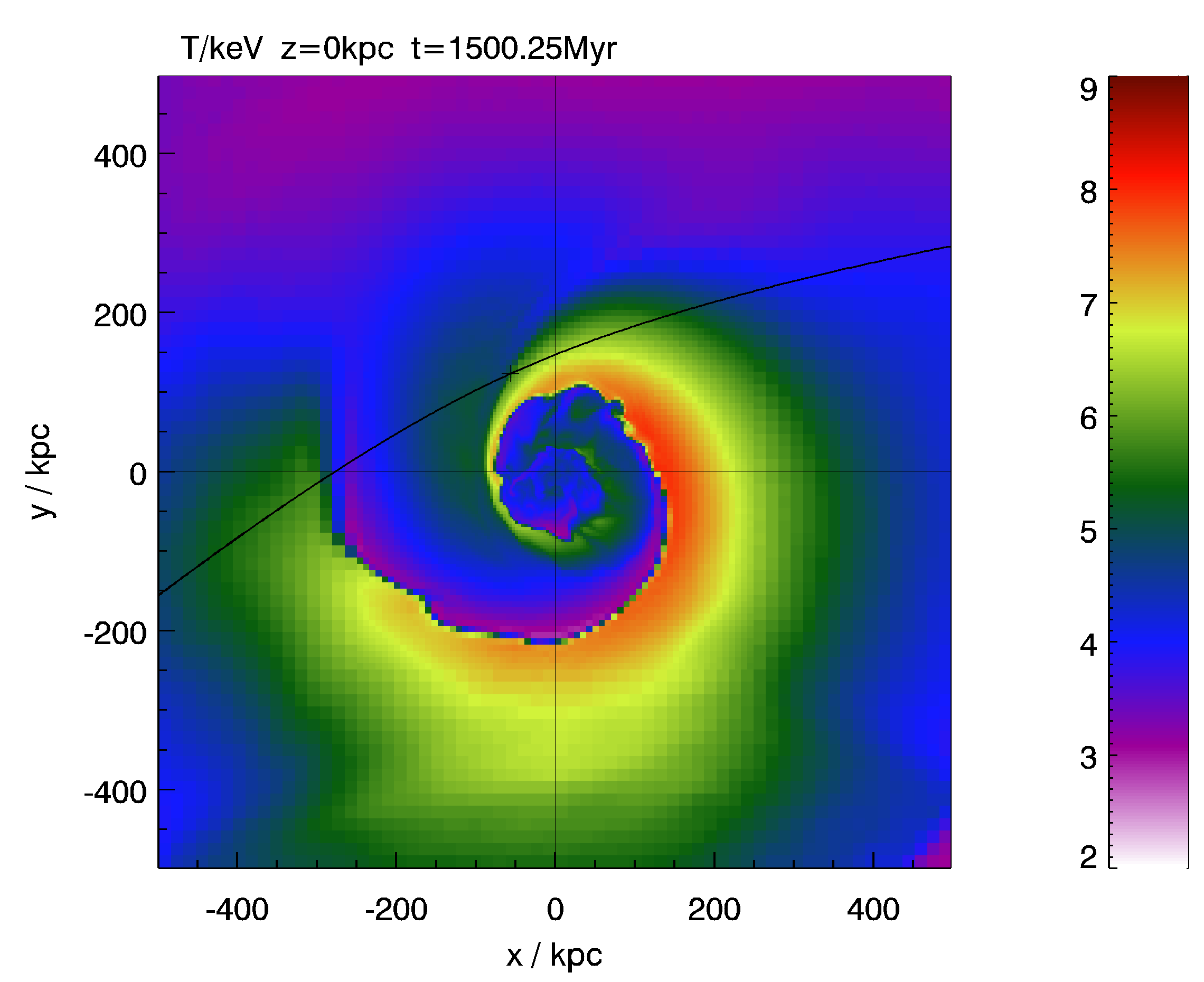}
\includegraphics[trim=300   00 0     0,clip,height=0.175\textheight]{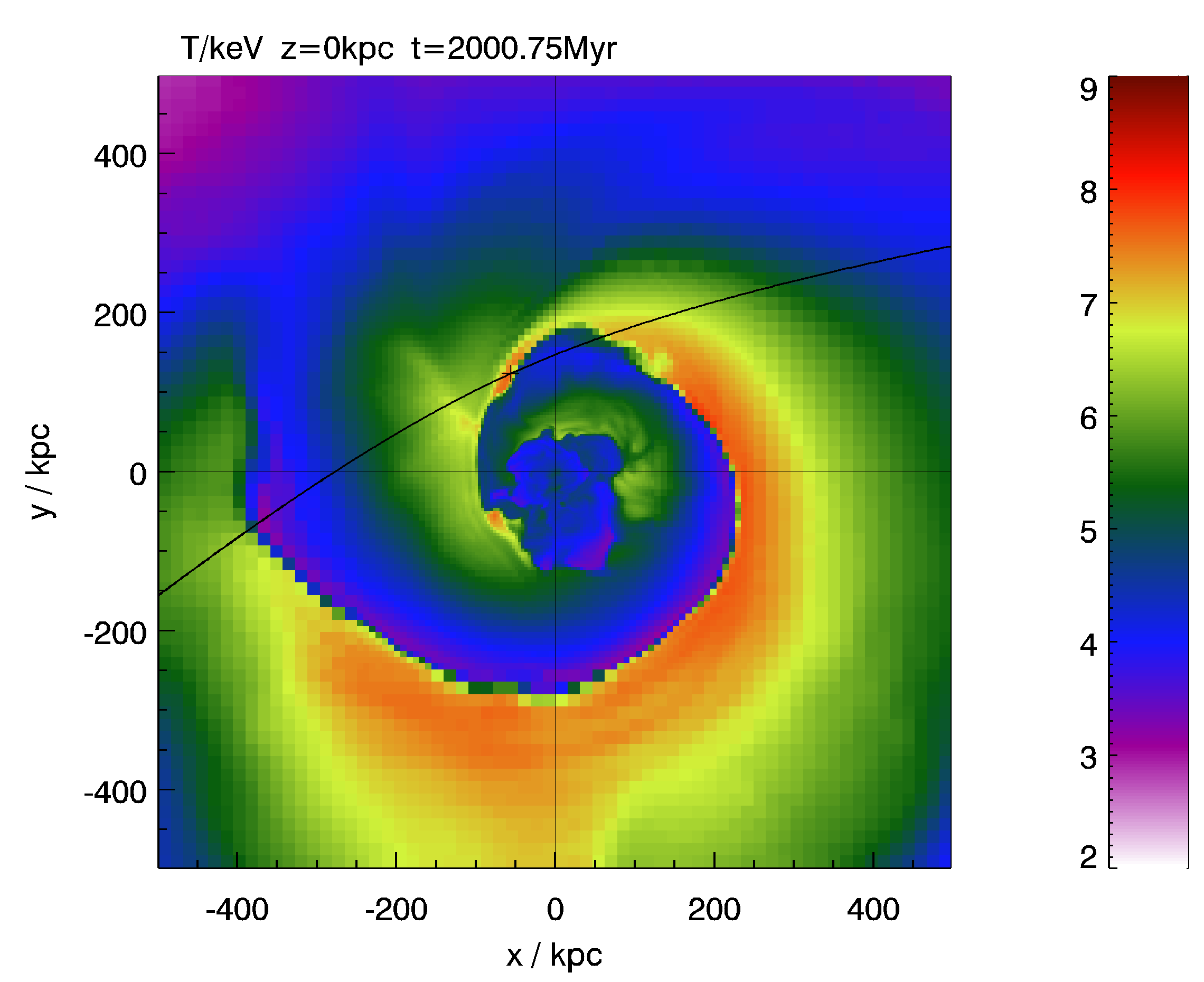}
\newline
\caption{Snapshots of the temperature in the orbital plane for the merger with mass ratio 2. We show the hydro+N-body results (upper row) with a delay of 250 Myr compared to the rigid potential run (bottom row). This hydro+N-body simulation has a low resolution of  10 kpc, which is at least partially responsible for the loss of the cool dense centre. The major cold front in the SW is still reproduced well, but the general distortion of the cluster by this major merger is not captured anymore by the rigid potential approximation.}
\label{fig:slicetemp_R2}
\end{figure*}
%
\begin{figure*}
\includegraphics[angle=0,width=0.8\textwidth]{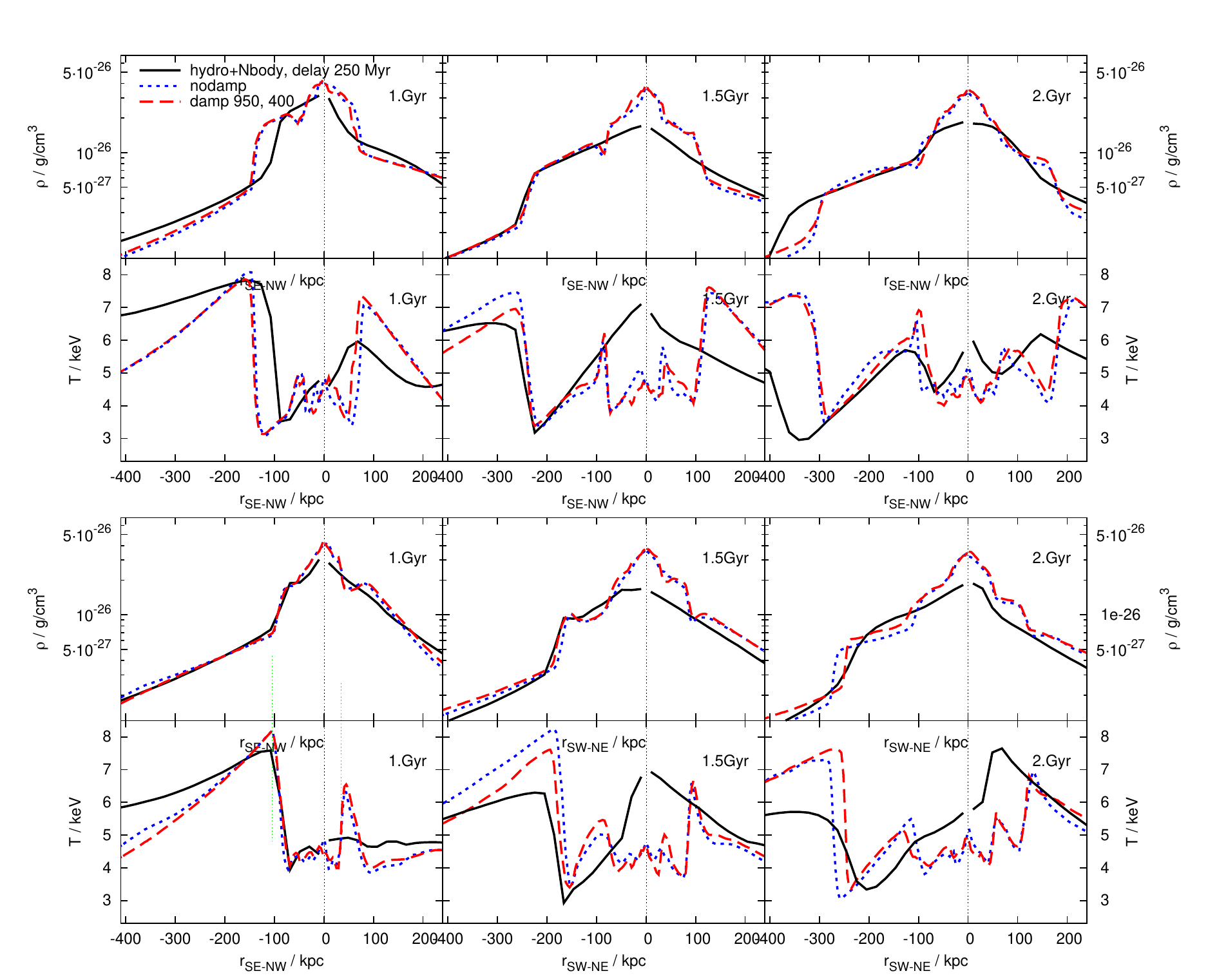}
\caption{%
Comparison of density and temperature profiles along the diagonals in the $xy$-plane. Same as Fig.~\ref{fig:compare_evolprofs_shift} but for a mass ratio of 2 between the clusters, which is already a major merger. The hydro+N-body simulation is plotted with a delay of $250\Myr$.}
\label{fig:compare_evolprofs_shift_R2}
\end{figure*}
%
%
A merger with a mass ratio of 2 is no minor merger anymore, and we perform this simulation with the purpose of exploring the limits of the RPA. Here, the subcluster has a scale radius of 416 kpc and we use a damping setting of $(R\damp/\Kpc,L\damp/\Kpc)=(950, 400)$. In Fig.~\ref{fig:slicetemp_R2} we compare temperature slices, in Fig.~\ref{fig:compare_evolprofs_shift_R2} profiles in the orbital plane along the diagonal directions at different timesteps. For this case, we find a delay of 250 Myr between RPA and hydro+N-body ones. 

In both methods, the central ICM starts sloshing in the cluster centre, and a clear primary CF is formed in the SW. However, in the hydro+N-body code the cool central gas moves completely out of the cluster centre, forming a CF towards the NE of the centre only at very late stages. In the RPA, the cool gas core is never completely displaced from the potential minimum. The comparison of profiles in  Fig.~\ref{fig:compare_evolprofs_shift_R2} reveals that the RPA still achieves a good agreement for the primary CF in the SW and SE, despite the nearly equal masses of both clusters. 

Thus, even for this major merger, the RPA recovers the evolution of the major CF. However, it comes to its limit regarding the morphology, the other CFs, and the large-scale distortion of the main cluster.

\section{Discussion and Summary}
%
\subsection{Origin of the delayed evolution in the RPA} \label{sec:discuss_delay}
%
\begin{figure*}
\includegraphics[angle=0,width=0.75\textwidth]{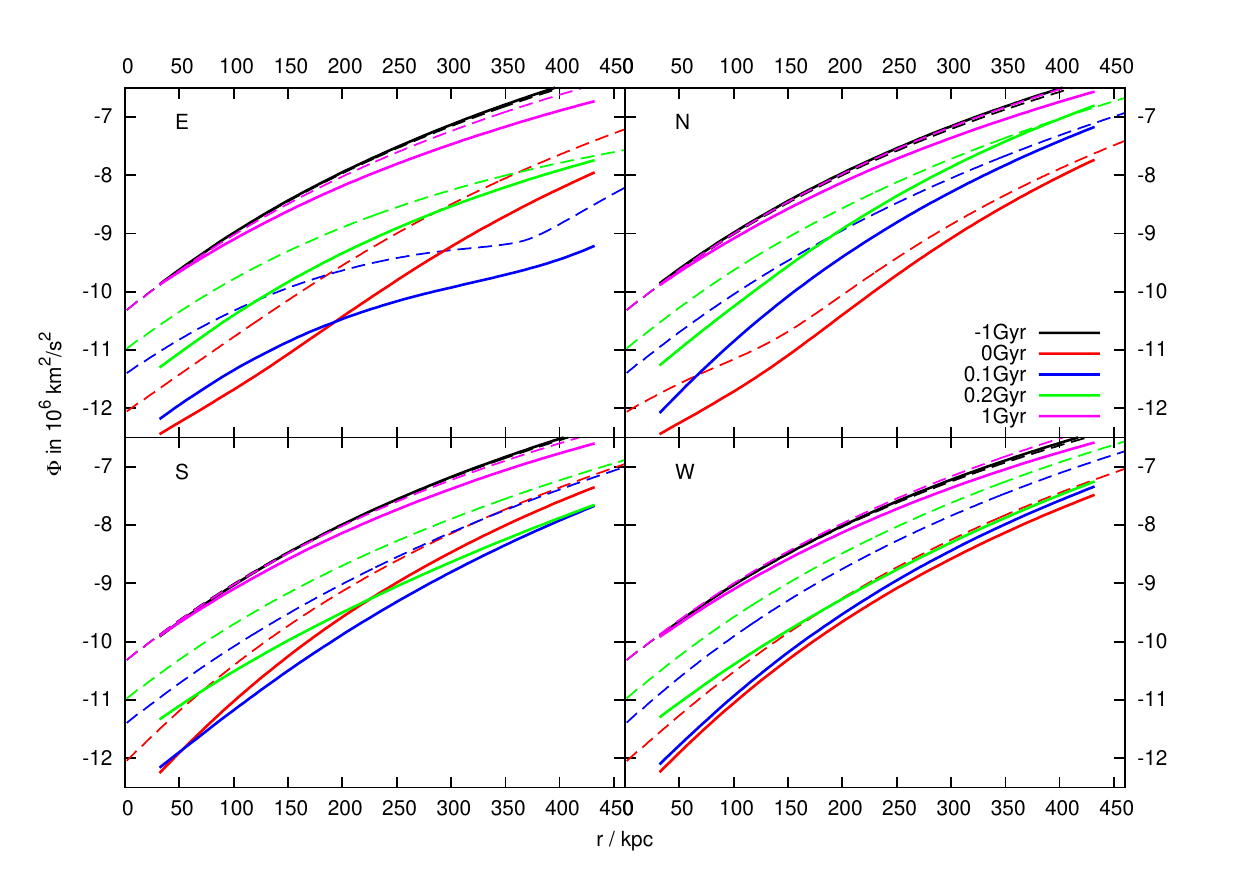}
\caption{Evolution of radial potential profiles in the main cluster towards N, W, S, E. The profiles are averaged azimuthally over $\pm 5\degree$.  Hydro+N-body simulations are shown by solid lines (thick lines in print version), rigid potential simulations by dashed lines (thin lines in print version). The colours (line styles in print version) code different timesteps, see legend. In both methods, the subcluster passage leads to a temporary deepening of the central potential. Additionally, the hydro+N-body method captures the tidal compression of the main cluster centre, leading to a stronger deepening and steepening of the potential. Both effects wear off with time and 1 Gyr after later the initial potential is nearly recovered. However, already after about 0.25 Gyr, the slopes of the potential are very similar in both methods.}
\label{fig:potential}
\end{figure*}
%
We have shown that the rigid potential approximation can reproduce the characteristics of gas sloshing well except for two artefacts, the higher temperatures at the outside of the CFs and the temporal lag in evolution.  We have traced back the former effect to the necessarily unrealistic gas motions in the outer cluster region in this approximation. In order to investigate the origin of the latter, we compare the evolution of radial potential profiles in the fiducial case in Fig.~\ref{fig:potential}. We show profiles towards the N, S, W and E, where we plot one direction per panel. Different timesteps are colour-coded (coded by line style in print version). The results from  hydro+N-body and from the RPA are shown by solid lines and dashed lines, respectively (thick and thin lines in print version). 

Generally, the central potential deepens and steepens during the passage of the subcluster. The direct overlap with the subcluster potential can also lead to a temporary local flattening of the potential, e.g.~at $t=0.1\Gyr$ in the E. In the RPA, the effect is solely due to the overlap of both potentials. The hydro+N-body code additionally captures the tidal compression of the cluster centre, which leads to the extra deepening and steepening during the pericentre passage evident from Fig.~\ref{fig:potential}. This leads to a temporarily different evolution of the central potentials between both methods. At 1 Gyr after pericentre passage, in both methods the potential is nearly back to its initial state, where the closest "recovery" is achieved in the S. This explains why the major, southern CF evolves so similar in both methods. 

\begin{figure*}
\includegraphics[angle=0,width=\textwidth]{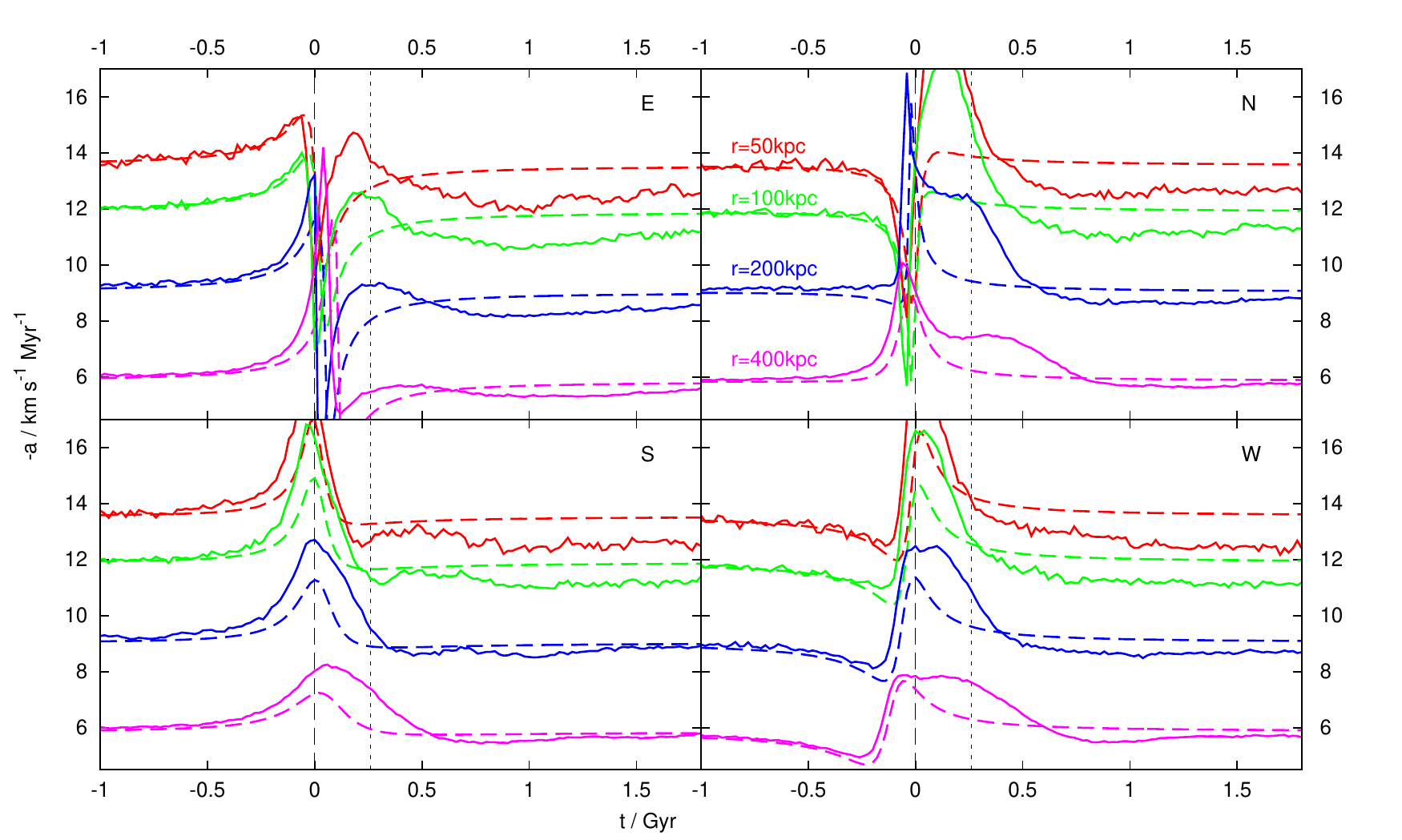}
\caption{Evolution of radial gravitational acceleration at different radii towards N, W, S, E. Comparison between hydro+N-body simulations (solid lines) and  rigid potential simulations (dashed lines). The colours code different radii, see labels. The effect of the subcluster passage is clearly seen in both methods. The effect is stronger in the hydro+N-body method because it captures the tidal compression of the cluster core. After about 0.25 to 0.5 Gyr, the acceleration is again similar in both methods. Only in the inner 100 kpc, the acceleration is slightly lower in the hydro+N-body method. Thus, after a slightly different onset, the sloshing proceeds in a very similar way afterwards.}
\label{fig:agrav}
\end{figure*}

In order to access the timescales of the potential evolution more directly, we study the evolution of the radial gravitational acceleration in different positions. Along each direction, we calculate the gravitational acceleration at four radii and plot their temporal evolution in Fig.~\ref{fig:agrav}. 
%
%
Again, there is one panel per direction, and the radii are colour-coded. Solid lines are for the hydro+N-body code, dashed for the RPA. Here we see more clearly the modification of the gravitational field described above. Prior to pericentre passage, the accelerations in both methods agree well. During pericentre passage, the tidal compression leads to stronger accelerations in the hydro+N-body method. After about 0.25 to 0.5 Gyr, the accelerations have reached the final levels at which they remain until the next pericentre passage. Outside 200 kpc, the initial acceleration is nearly recovered, whereas inside 100 kpc the final acceleration is slightly lower than the original one. 
The similarity of the accelerations, i.e.~the potential slopes, after the onset of sloshing accounts for the similar evolution of the CFs in both methods. The period of different potential evolution at and shortly after the pericentre passage is responsible for the different evolution during this phase and sends the hydro+N-body method ahead, because in a steeper potential the sloshing oscillations tend to be faster.

\subsection{Choice of damping parameters }
In Sect.~\ref{sec:method_damping} we suggested that the damping radius, $R\damp$, outside which the inertial frame correction will be switched off,  should be comparable to the diameter of the subcluster, i.e.~twice the subcluster scale radius. We investigated the impact of the damping parameters (Sect.~\ref{sec:fiducial_damping}) for all cases studied here and found that this choice leads indeed to the best possible reproduction of  all features. Moreover, we find that the evolution of CF radii is the same with this mild damping and in the undamped case, but CF radii are somewhat lower if the damping is chosen too strong.  This difference in behaviour can be used to confirm the choice of $R\damp$ for clusters where no hydro+Nbody simulation is yet available. As long as the CF radii in the damped simulation agree with the ones in the undamped case, the choice of $R\damp$ is reasonable. Regarding the damping scale length,  $L\damp$, we found that our results are not sensitive to this parameter,  and we recommend using $L\damp \approx 0.5R\damp$, as we have done here.  We also followed this strategy for choosing $R\damp$ and $L\damp$ in our simulations of Virgo and A496 (R11a,b).

\subsection{Summary - reliability of the rigid potential approximation}
We investigated the reliability of the rigid potential approximation described in Sect.~\ref{sec:inertframe} for simulations of minor merger induced gas sloshing. We use the hydro+N-body simulations of Z10 of the same scenario as the reference. Those capture the full evolution of the ICM and  DM components including dynamical friction, tidal compression and tidal stripping. In contrast, the rigid potential approximation treats the  potentials of the individual clusters as static and models only their relative motion. This simplification makes the rigid potential simulations faster by about a factor of 5 for the resolution used here, and more for higher resolution due to the imperfect scaling of the Poisson solver. This speed-up is very useful in several circumstances. For example, constraining the merger history of a given cluster by reproducing the observed sloshing signatures usually requires a large set of simulations (see R11a for Virgo cluster and R11b for Abell 496). Also  investigations of the impact of more time-consuming physics like viscosity (Z10) and thermal conduction benefit from a fast method for the basic process. 

While we expect and find temporal differences in the evolution of the gravitational potentials, regarding the gas sloshing, our main interest lies in the evolution of the ICM. Hence, we have simulated three representative merger scenarios using both methods and compared them in detail. We have shown that, except for two (correctable) artefacts, the rigid potential approximation reproduces the results of the hydro+N-body runs very well:
\begin{enumerate}
\item The rigid potential approximation reproduces the typical sloshing cold fronts and central cold spiral structure in, both,  morphology and orientation (Figs.~\ref{fig:slicetemp_500kpc}, \ref{fig:slicetemp_R20}, \ref{fig:slicetemp_R2}). 
\item The minor merger induces a characteristic large-scale asymmetry in the main cluster's ICM beyond the central cool spiral structure, which is reproduced at least qualitatively in, both, density and temperature (Fig.~\ref{fig:compare_evolprofs_shift}).
\item The outward motion of the cold fronts with time is delayed in the rigid potential approximation by typically 200 Myr compared to the hydro+N-body one. This originates from a temporarily different potential shape during pericentre passage because the rigid potential approximation does not capture tidal compression. The significantly different evolution is, however, short-lived, and the evolution proceeds very similar afterwards. Therefore, if this delay is corrected for, also the outward motion of the cold fronts is reproduced very well by the rigid potential approximation (Figs.~\ref{fig:evol_CFradii}, \ref{fig:evol_CFradii_shift_R20}). 
\item The ICM density on both sides of the cold fronts is reproduced well, and so is the temperature  inside the cold fronts (Figs.~\ref{fig:evol_CFradii_Temp}, \ref{fig:evol_CFradii_dens}, \ref{fig:evol_CFradii_Temp_R20}). The temperature at the outside of the cold fronts tends to be slightly too high compared to the reference simulations due to a necessarily unrealistic velocity field in the cluster outskirts in the rigid potential approximation (Sect.~\ref{sec:fiducial_damping}).  This effect is strongest if only the basic inertial frame correction is used (Sect.~\ref{sec:inertframe_basic}), and is milder for the improved version, where the inertial frame correction is only applied to the central part of the galaxy cluster and gradually dampened  towards large radii (Sect.~\ref{sec:method_damping}). Too strong dampening, however, causes  slightly too small cold front radii towards the directions aligned with the subcluster orbit. Our tests recommend dampening outside cluster-centric radii of about twice the subcluster scale radius, with a characteristic fall-off scale length comparable to the subcluster scale radius.
\end{enumerate}
Thus, the rigid potential approximation method can be employed e.g.~in order to disentangle the merger history for an observed cluster (R11a,b). The agreements in  items (i) and (ii) guarantee that the orientation of the subcluster orbit can be inferred correctly. Item (iii) ensures that the age of the cold front can be estimated reasonably, although here the uncorrected rigid potential approximation will over-estimate the age by about 200 Myr.  
If the age needs to be determined more accurately, especially during the onset of sloshing or the early evolution of the cold fronts, hydro+N-body simulations are required.
The mass of the subcluster can in principle be decoded from the contrast of density and temperature across the cold fronts. However, this attempt is intrinsically difficult, because usually the cold fronts are found within the cool cores of their host clusters, where the general gradient of, both, the density and temperature is in the same direction as the cold front discontinuity itself. Given that azimuthal and radial averaging in deriving radial profiles smears out the cold fronts over a certain radial extent, the contrast of all quantities between the inner and outer edge of a given cold front  includes the general variation of the quantity across in this radial range. The contributions of the general profile and the cold front discontinuity to this contrast are hard to separate, no matter which simulation method has been used.

\section*{Acknowledgments}
ER is  supported by the Priority Programme "Witnesses of Cosmic History'' of the DFG (German Research Foundation) and the supercomputing  grants NIC 3711 and 4368 at the John-Neumann Institut at the Forschungszentrum J\"ulich. JAZ is supported under the NASA postdoctoral program. We thank Marcus Br\"uggen for helpful discussions, and the referee Max Ruffert for his clarifying comments. The results presented were produced using the FLASH code, a product  of the DOE ASC/Alliances-funded Center for Astrophysical Thermonuclear Flashes  at the University of Chicago.

%
\bibliographystyle{mn2e}
\bibliography{library}

\begin{thebibliography}{}

\bibitem[\protect\citeauthoryear{Ascasibar \& Markevitch}{Ascasibar \&
  Markevitch}{2006}]{Ascasibar2006}
Ascasibar Y.,  Markevitch M.,  2006, ApJ, 650, 102

\bibitem[\protect\citeauthoryear{Bourdin \& Mazzotta}{Bourdin \&
  Mazzotta}{2008}]{Bourdin2008}
Bourdin H.,  Mazzotta P.,  2008, A\&A, 479, 307

\bibitem[\protect\citeauthoryear{Churazov, Forman, Jones \& Bohringer}{Churazov
  et~al.}{2003}]{Churazov2003}
Churazov E.,  Forman W.~R.,  Jones C.,    Bohringer H.,  2003, ApJ, 590, 225

\bibitem[\protect\citeauthoryear{Clarke, Blanton \& Sarazin}{Clarke
  et~al.}{2004}]{Clarke2004}
Clarke T.~E.,  Blanton E.~L.,    Sarazin C.~L.,  2004, ApJ, 616, 178

\bibitem[\protect\citeauthoryear{Dubey, Antypas, Ganapathy, Reid, Riley,
  Sheeler, Siegel \& Weide}{Dubey et~al.}{2009}]{Dubey2009}
Dubey A.,  Antypas K.,  Ganapathy M.~K.,  Reid L.~B.,  Riley K.,  Sheeler D.,
  Siegel A.,    Weide K.,  2009, Parallel Computing, 35, 512

\bibitem[\protect\citeauthoryear{Dupke, {White III} \& Bregman}{Dupke
  et~al.}{2007}]{Dupke2007}
Dupke R.,  {White III} R.~E.,    Bregman J.~N.,  2007, ApJ, 671, 181

\bibitem[\protect\citeauthoryear{Fabian, Sanders, Taylor \& Allen}{Fabian
  et~al.}{2005}]{Fabian2005centaurus}
Fabian A.~C.,  Sanders J.~S.,  Taylor G.~B.,    Allen S.~W.,  2005, MNRAS, 360,
  L20

\bibitem[\protect\citeauthoryear{Ghizzardi, Rossetti \& Molendi}{Ghizzardi
  et~al.}{2010}]{Ghizzardi2010}
Ghizzardi S.,  Rossetti M.,    Molendi S.,  2010, A\&A, 516, A32

\bibitem[\protect\citeauthoryear{Hernquist}{Hernquist}{1990}]{Hernquist1990}
Hernquist L.,  1990, ApJ, 356, 359

\bibitem[\protect\citeauthoryear{Johnson, ZuHone, Jones, Forman \&
  Markevitch}{Johnson et~al.}{2011}]{Johnson2011}
Johnson R.~E.,  ZuHone J.~A.,  Jones C.,  Forman W.,    Markevitch M.,  2011,
  eprint arXiv:1106.3489

\bibitem[\protect\citeauthoryear{Markevitch, Gonzalez, David, Vikhlinin,
  Murray, Forman, Jones \& Tucker}{Markevitch et~al.}{2002}]{Markevitch2002}
Markevitch M.,  Gonzalez A.~H.,  David L.,  Vikhlinin A.,  Murray S.,  Forman
  W.~R.,  Jones C.,    Tucker W.,  2002, ApJ, 567, L27

\bibitem[\protect\citeauthoryear{Markevitch, Ponman, Nulsen, Bautz, Burke,
  David, Davis, Donnelly, Forman, Jones, Kaastra, Kellogg, Kim, Kolodziejczak,
  Mazzotta, Pagliaro, Patel, {Van Speybroeck}, Vikhlinin, Vrtilek, Wise \&
  Zhao}{Markevitch et~al.}{2000}]{Markevitch2000}
Markevitch M.,  Ponman T.~J.,  Nulsen P. E.~J.,  Bautz M.~W.,  Burke D.~J.,
  David L.~P.,  Davis D.,  Donnelly R.~H.,  Forman W.~R.,  Jones C.,  Kaastra
  J.,  Kellogg E.,  Kim D.,  Kolodziejczak J.,  Mazzotta P.,  Pagliaro A.,
  Patel S.,  {Van Speybroeck} L.,  Vikhlinin A.,  Vrtilek J.,  Wise M.,    Zhao
  P.,  2000, ApJ, 541, 542

\bibitem[\protect\citeauthoryear{Markevitch \& Vikhlinin}{Markevitch \&
  Vikhlinin}{2007}]{Markevitch2007}
Markevitch M.,  Vikhlinin A.,  2007, Physics Reports, 443, 1

\bibitem[\protect\citeauthoryear{Markevitch, Vikhlinin \& Mazzotta}{Markevitch
  et~al.}{2001}]{Markevitch2001}
Markevitch M.,  Vikhlinin A.,    Mazzotta P.,  2001, ApJ, 562, L153

\bibitem[\protect\citeauthoryear{Mazzotta, Edge \& Markevitch}{Mazzotta
  et~al.}{2003}]{Mazzotta2003}
Mazzotta P.,  Edge A.~C.,    Markevitch M.,  2003, ApJ, 596, 190

\bibitem[\protect\citeauthoryear{Mazzotta \& Giacintucci}{Mazzotta \&
  Giacintucci}{2008}]{Mazzotta2008}
Mazzotta P.,  Giacintucci S.,  2008, ApJ, 675, L9

\bibitem[\protect\citeauthoryear{Mazzotta, Markevitch, Vikhlinin, Forman, David
  \& VanSpeybroeck}{Mazzotta et~al.}{2001}]{Mazzotta2001}
Mazzotta P.,  Markevitch M.,  Vikhlinin A.,  Forman W.~R.,  David L.~P.,
  VanSpeybroeck L.,  2001, ApJ, 555, 205

\bibitem[\protect\citeauthoryear{Million \& Allen}{Million \&
  Allen}{2009}]{Million2009sample}
Million E.~T.,  Allen S.~W.,  2009, MNRAS, 399, 1307

\bibitem[\protect\citeauthoryear{Owers, Nulsen, Couch \& Markevitch}{Owers
  et~al.}{2009}]{Owers2009hifid}
Owers M.~S.,  Nulsen P. E.~J.,  Couch W.~J.,    Markevitch M.,  2009, ApJ, 704,
  1349

\bibitem[\protect\citeauthoryear{Roediger, Br\"{u}ggen, Simionescu,
  B\"{o}hringer, Churazov \& Forman}{Roediger et~al.}{2011}]{Roediger2011}
Roediger E.,  Br\"{u}ggen M.,  Simionescu A.,  B\"{o}hringer H.,  Churazov E.,
    Forman W.~R.,  2011, MNRAS, 413, 2057

\bibitem[\protect\citeauthoryear{Roediger, Lovisari, Dupke, Ghizzardi \&
  Br\"{u}ggen}{Roediger et~al.}{2011}]{Roediger2011a496}
Roediger E.,  Lovisari L.,  Dupke R.~A.,  Ghizzardi S.,    Br\"{u}ggen M.,
  2011, MNRAS, submitted

\bibitem[\protect\citeauthoryear{Sanders \& Fabian}{Sanders \&
  Fabian}{2006}]{Sanders2006centaurus}
Sanders J.~S.,  Fabian A.~C.,  2006, MNRAS, 371, 1483

\bibitem[\protect\citeauthoryear{Sanders, Fabian \& Dunn}{Sanders
  et~al.}{2005}]{Sanders2005perseus}
Sanders J.~S.,  Fabian A.~C.,    Dunn R. J.~H.,  2005, MNRAS, 360, 133

\bibitem[\protect\citeauthoryear{Sanders, Fabian \& Taylor}{Sanders
  et~al.}{2009}]{Sanders2009_2a}
Sanders J.~S.,  Fabian A.~C.,    Taylor G.~B.,  2009, MNRAS, 396, 1449

\bibitem[\protect\citeauthoryear{Simionescu, Werner, Forman, Miller, Takei,
  B\"{o}hringer, Churazov \& Nulsen}{Simionescu et~al.}{2010}]{Simionescu2010}
Simionescu A.,  Werner N.,  Forman W.~R.,  Miller E.~D.,  Takei Y.,
  B\"{o}hringer H.,  Churazov E.,    Nulsen P. E.~J.,  2010, MNRAS, 405, 91

\bibitem[\protect\citeauthoryear{Taylor \& Babul}{Taylor \&
  Babul}{2001}]{Taylor2001}
Taylor J.~E.,  Babul A.,  2001, ApJ, 559, 716

\bibitem[\protect\citeauthoryear{Vikhlinin, Markevitch \& Murray}{Vikhlinin
  et~al.}{2001}]{Vikhlinin2001}
Vikhlinin A.,  Markevitch M.,    Murray S.~S.,  2001, ApJ, 551, 160

\bibitem[\protect\citeauthoryear{ZuHone, Markevitch \& Johnson}{ZuHone
  et~al.}{2010}]{ZuHone2010}
ZuHone J.~A.,  Markevitch M.,    Johnson R.~E.,  2010, ApJ, 717, 908

\bibitem[\protect\citeauthoryear{ZuHone, Markevitch \& Lee}{ZuHone
  et~al.}{2011}]{ZuHone2011}
ZuHone J.~A.,  Markevitch M.,    Lee D.,  2011, eprint arXiv:1108.4427

\end{thebibliography}

\bsp

\label{lastpage}

\end{document}